\documentclass[useAMS,usenatbib]{mn2e}
\usepackage{times}
\usepackage{natbib}
\usepackage[dvips]{graphicx}
\usepackage{amsmath}
\newcommand{\chandra}{{\it Chandra\/}}
\newcommand{\apjs}{ApJS}
\newcommand{\apj}{ApJ}
\newcommand{\aap}{A{\&}A}
\newcommand{\mnras}{MNRAS}
\newcommand{\aj}{AJ}
\newcommand{\araa}{ARA\&A}
\newcommand{\nar}{New A Rev.}
\newcommand{\apjl}{ApJ}
\newcommand{\nat}{Nature}

\newcommand{\mflux}{{erg~cm$^{-2}$~s$^{-1}$~Hz$^{-1}$}}
\newcommand{\lum}{{erg~s$^{-1}$}}
\newcommand{\mlum}{{erg~s$^{-1}$~Hz$^{-1}$}}
\newcommand{\asca}{{\it ASCA\/}}
\newcommand\ion[2]{#1$\;${\scshape{#2}}}

\begin{document}

\title[High-Luminosity Disc-Like Emitters]{X-ray and Multiwavelength
Insights into the Inner Structure of 
High-Luminosity Disc-Like Emitters}

\author[B.~Luo et al.]{
B.~Luo,$^{1,2}$\thanks{E-mail: lbin@astro.psu.edu}
W.~N.~Brandt,$^{1,2}$
M.~Eracleous,$^{1,2}$
Jian~Wu,$^{1,3}$
P.~B.~Hall,$^{4}$
A.~Rafiee,$^{5}$
\newauthor
D.~P.~Schneider,$^{1,2}$
and Jianfeng~Wu$^{1,2}$\\
$^{1}${Department of Astronomy \& Astrophysics, 525 Davey Lab,
The Pennsylvania State University, University Park, PA 16802, USA}\\
$^{2}${Institute for Gravitation and the Cosmos, 
The Pennsylvania State University, University Park, PA 16802, USA}\\
$^{3}${College of Information Sciences and Technology, 
316F IST Building, Pennsylvania State University, 
University Park, PA 16802, USA}\\
$^{4}${Department of Physics \& Astronomy, York University, 4700 Keele Street, Toronto, ON M3J 1P3, Canada}\\
$^{5}${Department of Physics, Astronomy, and Geosciences, Towson University, Towson, MD 21252, USA}}
\date{ }
\maketitle
\begin{abstract}
We present X-ray and multiwavelength studies of a 
sample of eight high-luminosity
active galactic nuclei (AGNs) with disc-like H$\beta$ emission-line profiles
selected from the Sloan Digital Sky Survey Data Release
7.
These sources have higher redshift ($z\approx0.6$) 
than the majority of the known disc-like
emitters, and they occupy a
largely unexplored space in the luminosity-redshift plane.
Seven sources have typical AGN X-ray spectra with power-law photon indices of
$\Gamma\approx1.4$--2.0; two of them show some X-ray absorption
(column density $N_{\rm H}\approx10^{21}$--$10^{22}$~cm$^{-2}$ for neutral 
gas). 
The other source, J$0850+4451$,
has only three hard X-ray photons detected and is probably heavily obscured
($N_{\rm H}\ga3\times10^{23}$~cm$^{-2}$). This object 
is also identified 
as a low-ionization
broad absorption line (BAL) quasar based on
\ion{Mg}{ii} $\lambda2799$ absorption; it is the first disc-like
emitter reported that is also a BAL quasar.
The IR-to-UV spectral energy distributions (SEDs) 
of these eight sources are similar to the mean SEDs of
typical quasars with a UV ``bump'', suggestive of
standard accretion discs radiating with
high efficiency, which differs from low-luminosity disc-like emitters.
Studies of the \hbox{X-ray}-to-optical power-law slope parameters ($\alpha_{\rm OX}$)
indicate that there is no significant excess X-ray emission in 
these high-luminosity disc-like emitters.
Energy budget analysis suggests that for disc-like emitters in general, 
the inner disc
must illuminate and ionize the outer disc efficiently 
($\approx15\%$ of the nuclear ionizing radiation is required on average)
via
direct illumination and/or scattering. 
Warped accretion discs are probably needed for direct illumination to
work in high-luminosity objects, as their geometrically 
thin inner discs decrease the
amount of direct illumination possible for a flat disc.

\end{abstract}
\section{INTRODUCTION}

In a subset of active galactic nuclei (AGNs) the broad, optical 
emission lines (most notably the Balmer lines) have profiles that 
can be well described by models that attribute the emission 
to the surface of a relativistic, Keplerian accretion disc. 
The most striking of the disc-like profiles are double peaked and 
are reminiscent of the line profiles observed in cataclysmic 
variables 
(e.g., \citealt{Greenstein1959}; \citealt{Young1980};
\citealt*{Young1981}; \citealt{Marsh1988})
and attributed to their accretion discs.  Double-peaked lines 
in AGNs have received a great deal of attention since their discovery
(e.g., \citealt*{Stauffer1983}; \citealt{Oke1987})
and have motivated the development of disc emission models to describe 
their profiles (e.g., \citealt*{Chen1989a}; \citealt{Chen1989}; \citealt*{Rokaki1992}).
In these models, the emission originates in the photoionized skin of 
the disc at radii from a few hundred to a few thousand gravitational 
radii from the centre (the gravitational radius is defined as 
$R_{\rm G}=GM/c^2$). This region of the accretion disc has actually 
been suggested to be the broad-line region (BLR) in AGNs, or at least 
to be responsible for producing the low-ionization broad lines 
(see \citealt{Collin1987}; \citealt*{Eracleous2009}; 
\citealt{Gaskell2009} and references therein).
Although other mechanisms have been considered for the origin of 
double-peaked lines, emission from the outer accretion disc is the 
most plausible one (e.g., see arguments in \citealt{Eracleous2003}).

With an appropriate choice of model parameters, axisymmetric disc 
models can also describe profiles that are not obviously double peaked 
but instead have double shoulders that are not widely separated or 
flat tops (this can be accomplished by choosing a small inclination 
angle or a very large outer radius, for example; see 
\citealt{Eracleous2001,Eracleous2003} and references therein). 
Moreover, the initial line-profile models have been extended by 
introducing non-axisymmetric perturbations to the disc emissivity 
(e.g., an eccentricity or a spiral arm; see 
\citealt{Eracleous1995,Storchi2003} and references therein) 
to describe an even wider variety of profile shapes, such as those 
that are very asymmetric (e.g., with one dominant shoulder). 
Finally, the inclusion in the models of radiative-transfer effects 
in the disc atmosphere can result in profiles that are flat topped or 
even single peaked (see \citealt{Murray1997}; \citealt*{Flohic2012}). Here we 
consider all profiles that have double peaks, shoulders, asymmetries, 
or other features 
that resemble those from 
disc models (regardless whether the discs in these models 
are axisymmetric or not) as disc like. Adopting as our working hypothesis 
that the outer disc accounts for much of the low-ionization
BLR emission and that disc-like lines give us a 
direct view of this BLR, we use the lines as tools to gain insights 
into its structure and physical properties.

Disc-like emitters are usually identified via their 
H$\alpha$ or H$\beta$ lines 
\citep[e.g.,][]{Eracleous2003,Strateva2003,Shen2011}, and in a few cases, 
disc-like \ion{Mg}{ii} $\lambda2799$ lines\footnote{The \ion{Mg}{ii} line is actually a doublet at $\lambda2796$ and $\lambda2803$, and it
is treated as a single line throughout this paper.} have been observed 
\citep[e.g.,][]{Strateva2003,Eracleous2004,Luo2009}. 
The disc-like lines are among the broadest optical emission lines 
known in AGNs, with full widths at half maximum (FWHM) sometimes exceeding 
15\,000~km~s$^{-1}$ \citep[e.g.,][]{Strateva2008}.
The fraction of disc-like emitters among the 
general population of AGNs is $\approx3\%$, determined based on a 
survey of $\approx3000$
low-redshift ($z<0.33$) AGNs \citep{Strateva2003}\footnote{Only a fraction 
of the line profiles identified by \citet{Strateva2003} are clearly 
double peaked. The others have double or asymmetric shoulders or flat tops 
which are very suggestive of emission from an accretion disc. 
As demonstrated by those authors, the line asymmetries can explained 
well by disc models.}
selected from the Sloan Digital Sky Survey (SDSS; \citealt{York2000}); 
the fraction among radio-loud AGNs is probably higher by a factor of 
a few \citep{Eracleous1994,Eracleous2003}. Most known disc-like emitters
are located at $z\la0.5$, with the highest-redshift one, J$0331-2755$, 
having been discovered
serendipitously in the Extended \chandra\ Deep Field-South at $z=1.37$
\citep{Luo2009}.

After almost 30 years since their discovery, there are still two
fundamental open questions 
regarding the nature of disc-like emitters. 
The first one is the energy budget problem.
In many disc-like emitters, the gravitational power available
locally in the line-emitting region is insufficient to power the
observed lines, and it has been suggested that
the outer disc is photoionized by external illumination. 
The illumination mechanism, however, is not clear. For the initial sample of 
\hbox{low-luminosity} disc-like emitters, a central X-ray emitting
elevated structure was proposed to illuminate the outer disc
\citep[e.g.,][]{Chen1989a}. However, a significant number of 
high-luminosity disc-like emitters have been
discovered recently, which are not expected to possess such an elevated structure 
\citep[e.g.,][]{Strateva2008}. 
A scattering medium above the disc (e.g., an outflowing wind) can also
redirect continuum photons from the centre to the outer disc, as has 
been discussed in \citet{Dumont1990}. \citet{Cao2006} made a 
somewhat different suggestion that a jet is the scattering medium.
It is thus interesting to test 
whether there is any excess X-ray emission from high-luminosity disc-like
emitters that acts as the power source, or alternatively, whether
the radiation from the centre is transported to the outer disc via scattering.

The other open question regards the difference between
disc-like emitters and more typical AGNs. 
If the BLR is associated with the accretion disc, why do the
\hbox{disc-like} broad lines appear in only $\approx3\%$ of AGNs?
Under certain accretion-disc conditions, such as a nearly
face-on disc \citep[e.g.,][]{Corbin1995} or a disc-emission region 
with a large ratio of the outer to inner 
radius 
(e.g., \citealt{Dumont1990}; \citealt*{Jackson1991}),
the disc emission lines
could appear single peaked. However, these alone are probably not sufficient
to solve the problem (see \citealt{Eracleous2009} for a review).
One promising model that addresses this problem involves radiative
transfer in an accelerating disc wind:
a line from the disc that would have a disc-like profile in the
absence of a disc wind can appear single peaked
because of the non-axisymmetric emissivity pattern produced by
the radiative transfer \citep{Murray1997}.
This disc-wind model suggests that 
disc-like emitters with luminosities close to the
Eddington limit (likely high-luminosity AGNs) 
should be rare as the
radiatively driven winds are likely denser and would have prevented
the formation of disc-like lines.

Understanding the nature of disc-like emitters is 
especially valuable for painting a general picture of the 
inner structure of AGNs: the detailed inner components of AGNs,
including the accretion disc, disc wind, BLR, and scattering medium,
are closely related to the appearance of disc-like emission lines.
High-luminosity disc-like emitters appear to be one key
to constrain the possible models that can answer the unresolved questions.
Moreover, disc-like emitters have not been
studied in detail at higher redshifts ($z\ga0.5$), when the Universe 
was nearly half of its age and possible cosmic evolution over Gyrs 
could have altered their general properties.
We present in this paper
X-ray and multiwavelength studies of a sample of eight high-luminosity 
disc-like emitters that are also at relatively high redshifts ($z\approx0.6$).
In Section~2 we describe the selection procedure of the sample.
In Section~3 we present the line-emitting region parameters for these
sources derived from line-profile fitting. We study the X-ray and
multiwavelength properties in Sections~4 and 5, respectively.
In Section~6 we discuss the implications of our results for the 
energy budget problem, and for the first time, we consider 
carefully a variety of mechanisms that can balance the energy budget of the 
line-emitting region in high-luminosity disc-like emitters. We also discuss
the disc-wind model addressing the connection 
between disc-like emitters and more typical AGNs.
We summarize in Section~7.
Throughout this paper,
we adopt a cosmology with
$H_0=70.4$~km~s$^{-1}$~Mpc$^{-1}$, $\Omega_{\rm M}=0.272$,
and $\Omega_{\Lambda}=0.728$ \citep[e.g.,][]{Komatsu2011}.

\section{SAMPLE SELECTION} \label{sample}

We systematically searched for high-luminosity
disc-like
emitters from the SDSS Data Release 7 (DR7; \citealt{Abazajian2009}) 
quasar catalog \citep{Schneider2010}.
We defined a parent sample of 3132 quasars with 
$z=0.5$--0.8, $L_{\rm 2500~\AA}>10^{30}$~\mlum\ 
(2500~\AA\ monochromatic luminosity; \citealt{Shen2011}), and a minimum
continuum signal-to-noise ratio (around H$\beta$; unbinned) $\ge7$.
The redshift range was chosen so that we are 
exploring the relatively high-redshift space ($z>0.5$)
and
the H$\beta$ line is well covered by the SDSS spectrum ($z<0.8$).
We fit the SDSS spectra
using the spectral-fitting routine presented in \citet{Wu2009}, which
can perform continuum subtraction and \hbox{multi-component} Gaussian fitting
of emission lines.
The underlying
continuum consists of three components: a
power-law spectrum, the Balmer bump (between 2000 and 4000~\AA),
and the iron emission forest (modeled with 
templates from \citealt{Vestergaard2001} and
\citealt*{Veron2004}). After subtracting the continuum,
we used 3--4 Gaussian components to fit the
broad H$\beta$ line, along with 2--3 Gaussian components for each of the
[\ion{O}{iii}]~$\lambda$4960 and $\lambda$5007 narrow lines.
We visually inspected each spectrum that has its H$\beta$ line best fitting
by at least one blueshifted and one redshifted Gaussian profile,
and identified 14 secure disc-like emitters.
In this high-luminosity AGN sample, we did not find any source
having an H$\beta$ line with two well-defined and well-separated 
peaks as in those original 
double-peaked emitters studied in \citet{Eracleous1994}. 
All the H$\beta$ line profiles of these 14 objects have 
broad blue and red shoulders or flat tops that suggest disc-like emission (see 
Section 3 below for disc-model fits to the profiles).\footnote{Of course,
since we are verifying that a profile is disc like after we make an initial
selection, we cannot be sure that we have selected all disc-like profiles out
of a given sample. Nonetheless, the analysis presented in this study and its
conclusions do not depend on selecting all disc-like profiles out of the 
sample that we have searched.}
Of these 14 sources, 12 are radio quiet (radio-loudness parameter
$R<10$; see Section 5.1 below for definition) and two are radio loud ($R\ge10$).

We selected the six most luminous
radio-quiet sources among these 14 disc-like emitters and
obtained \chandra\ snapshot observations of them in Cycle 12;
we chose radio-quiet sources as their X-ray spectra are not dominated
by jet-linked emission \citep[e.g.,][]{Worrall1987,Miller2011}.
One other object among the 14 sources, 
J$1531+2420$, has an archival 
\chandra\ observation and
is thus included in this study.
J$1531+2420$ was targeted by \chandra\
as one of the {\it Advanced Satellite for Cosmology and Astrophysics} 
({\it ASCA}) SHEEP (Search for the High-Energy Extragalactic Population)
survey
sources \citep[e.g.,][]{Nandra2003}. 
Furthermore, we include in our sample J$2125-0813$, the luminous
disc-like emitter
presented in \citet{Strateva2008} that has been targeted in additional
\chandra\ and {\it XMM-Newton} observations. Therefore,
our sample consists of eight high-luminosity
($L_{\rm 2500~\AA}>10^{30}$ \mlum) and 
relatively high-redshift ($z\approx0.6$) disc-like emitters
that have \chandra\ coverage. 
Their $i$-band absolute AB magnitudes range from
$-24.8$ to $-27.2$; these sources are within the top $\approx5\%$ 
most-luminous quasars in the SDSS DR7 quasar 
catalog in the redshift range of 0.5--0.8.
The \chandra\ observations
of these sources are listed in Table~\ref{tbl-log}; the exposure times
range from 4~ks to 10~ks except for J$2125-0813$, 
which has a 40~ks \chandra\ exposure.
The redshifts for these sources are obtained from
\citet{Hewett2010}, which provides the best available redshift measurements 
(mainly from the [\ion{O}{iii}] lines for our objects) for 
SDSS quasars.

\begin{figure}
\centerline{
\includegraphics[scale=0.5]{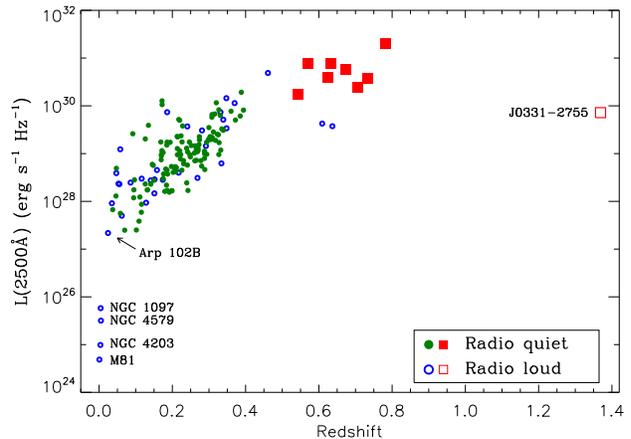}
}
\caption{
Redshift-luminosity distribution of known disc-like emitters
\citep[adapted from][]{Strateva2008}.
Green filled dots represent radio-quiet sources, 
blue open dots represent radio-loud
sources, red filled squares represent the eight radio-quiet objects 
studied here, and the red open square represents the highest-redshift
disc-like emitter, J$0331-2755$ \citep{Luo2009}.
The positions of the prototypical disc-like emitter, Arp 102B, and a few local
disc-like emitters are indicated. (A color version of this figure is
available in the online journal.)
\label{fig-lz}}
\end{figure}

In Figure~\ref{fig-lz}, we show the positions of our sample objects in the 
redshift versus 2500 \AA\ monochromatic-luminosity plane (see \citealt{Strateva2008} and references therein);
some known disc-like emitters, including the prototypical source
Arp 102B, are shown.
Our eight sources are all radio quiet, and they
sample a largely unexplored space in the luminosity-redshift plane.
Although these objects were selected based on their disc-like 
H$\beta$ line profiles, the \ion{Mg}{ii} $\lambda2799$ line is 
covered by the SDSS spectrum in all cases (see details below in
Section 3). We also discovered broad UV absorption features 
in J$0850+4451$, which are discussed in Section 4.2 below;
the other objects do not show apparent UV absorption in their spectra. 
The highest-redshift known
disc-like emitter, J$0331-2755$ \citep{Luo2009}, has a luminosity
($L_{\rm 2500~\AA}=7.2\times10^{29}$ \mlum) comparable to our sample objects,
and thus we include this source in some of our discussion
(Section~\ref{sec-discuss}).
We caution that J$0331-2755$ is radio loud, and some of its X-ray emission may
originate from a jet.

\section{LINE PROFILES AND DISK-MODEL FITS} \label{sec-spec}

\begin{figure*}
\centerline{
{\includegraphics[scale=0.5]{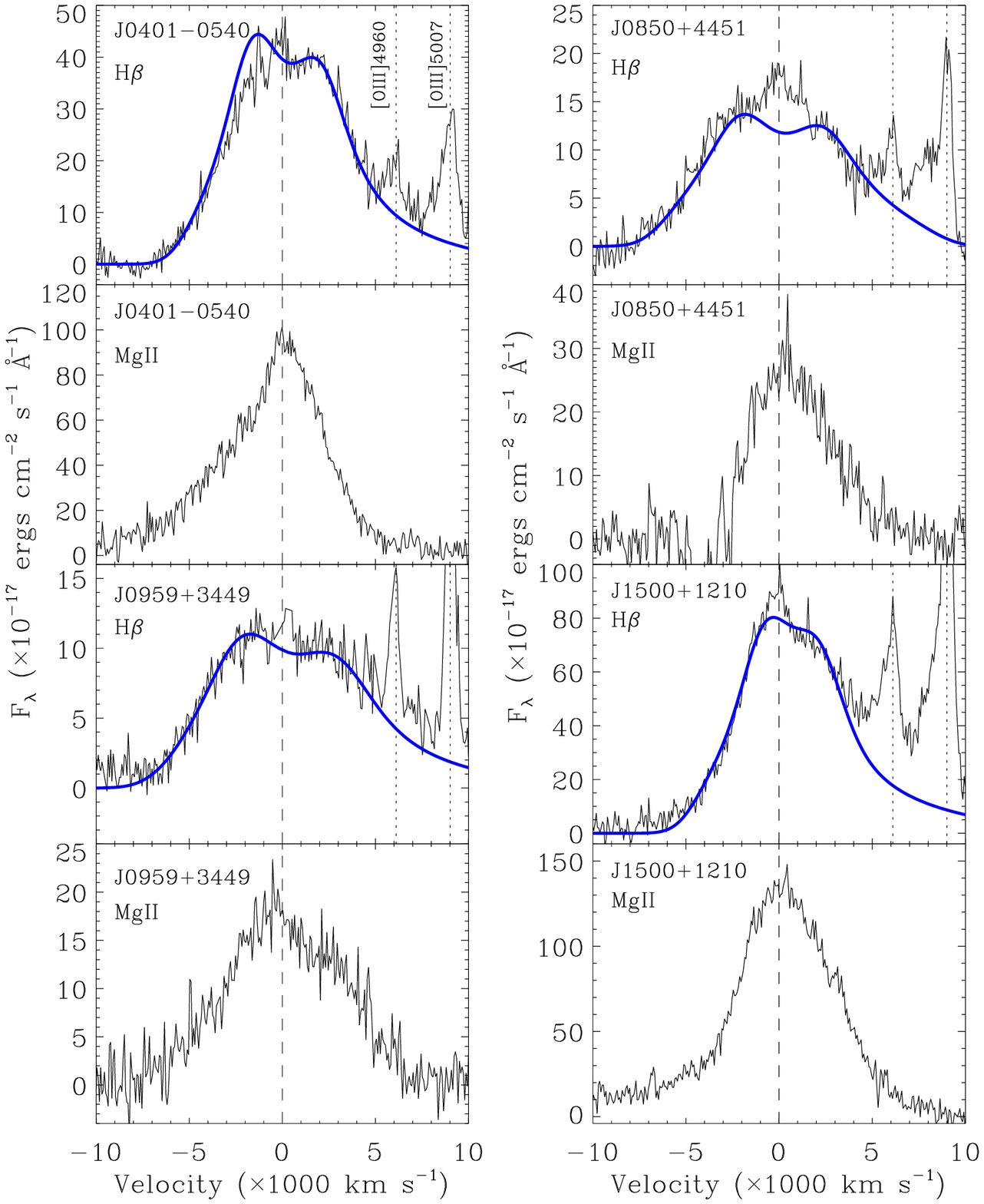}
\includegraphics[scale=0.5]{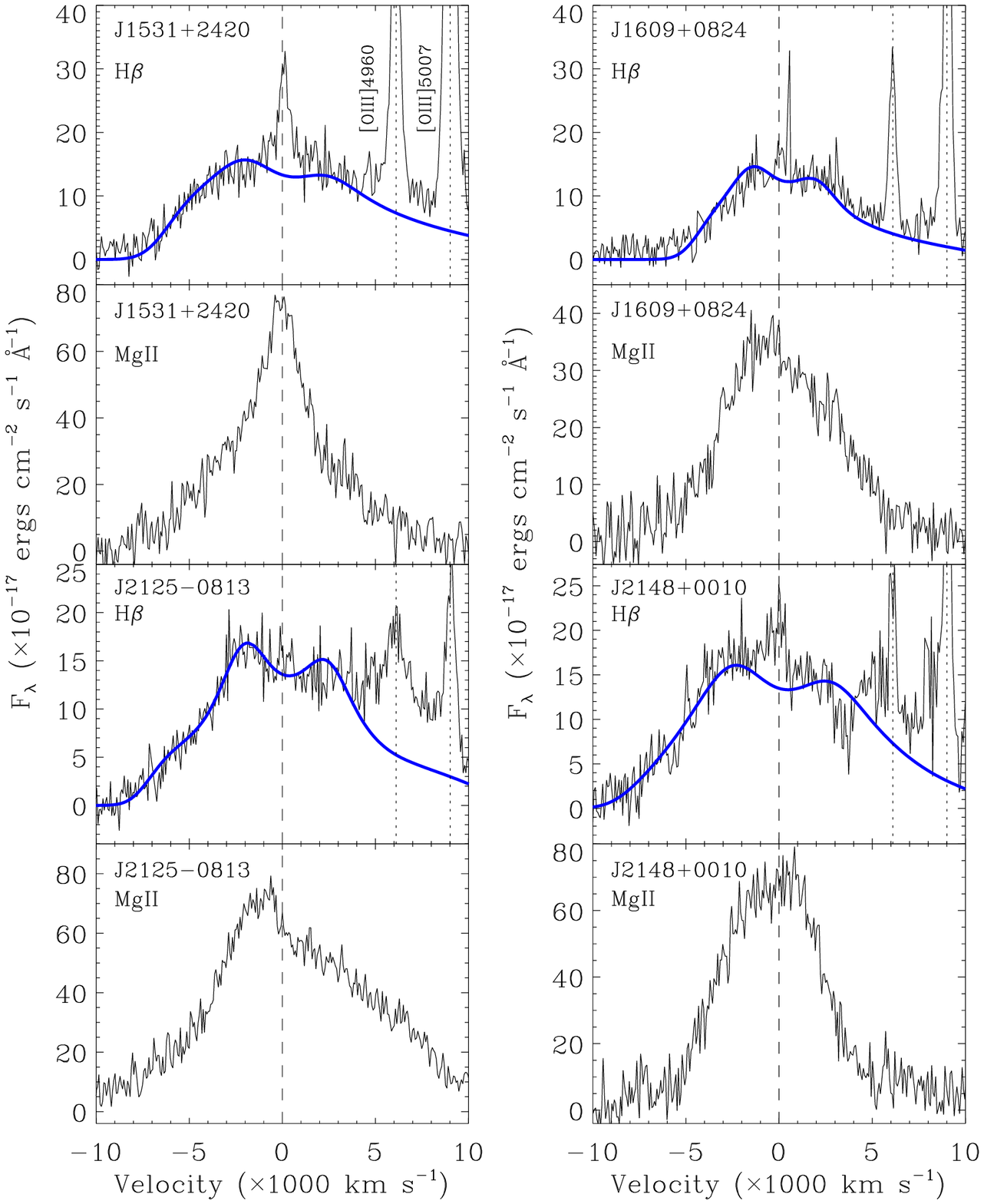}
}
}
\caption{
H$\beta$ and \ion{Mg}{ii} $\lambda2799$ line profiles of
the eight objects in the rest frame derived by subtracting the
power law spectrum and iron emission forest from the SDSS spectra
(see Section~\ref{sample}).
The H$\beta$ line profiles are overlaid with
the best-fitting
models (blue curves) of emission from a relativistic Keplerian disc.
(A color version of this figure is available in the online journal.)
\label{fig-opspec}}
\end{figure*}

The continuum-subtracted H$\beta$ line profiles for the eight objects
are displayed in 
Figure~\ref{fig-opspec}, derived from the spectral-fitting routine described
in Section~\ref{sample}. 
The FWHMs of the broad H$\beta$ lines, listed
in Table~\ref{tbl-fit}, are in the range of $\approx6000$--11000~km~s$^{-1}$;
for comparison, the median FWHM of H$\beta$ lines in general
AGNs with similar luminosities (from 
the \citealt{Shen2011} SDSS DR7 catalog) is only $\approx4000$~km~s$^{-1}$.
Since the Baldwin effect for H$\beta$ is not significant
in typical AGNs \citep[e.g.,][]{Dietrich2002}, we did not assess it here for the
disc-like emitters.

We fit the disc-like H$\beta$ line profiles (after continuum subtraction) 
with the relativistic Keplerian disc model of
\citet{Chen1989}. The model assumes that the line 
originates from the surface of a circular Keplerian disc;
the free parameters are the inner and outer radii of the emission regions
($R_{\rm in}$ and $R_{\rm out}$ in units of gravitational radii 
$R_{\rm G}$), the disc inclination angle ($i$),
the disc emissivity power-law index ($q_1$ for a simple power-law model 
or $q_1$, $q_2$, and 
$R_{\rm b}$ for a broken power-law model),\footnote{A simple 
power-law model ($\epsilon \propto R^{-q_1}$) was assumed first for the emissivity. If the wings of the 
line profile could not be well fit, we then adopted the more 
complex broken power-law model ($\epsilon \propto R^{-q_1}$
when $R<R_{\rm b}$ and $\epsilon \propto R^{-q_2}$
when $R\ge R_{\rm b}$).}
and the broadening parameter
($\sigma$). The central narrow/broad 
line was excluded from the fit if present.
The best-fitting parameters for the sample objects are reported 
in Table~\ref{tbl-fit},
and the model line profiles are superposed on the spectra 
in Figure~\ref{fig-opspec}.
The parameters are within the typical ranges found for  
other disc-like emitters \citep[e.g.,][]{Eracleous1994,Eracleous2003,Strateva2003}.
These accretion discs generally have small inclination angles ($i<35\degr$,
relatively face on).
The inner radius of the emission region 
is a crucial parameter for determining the available local viscous energy 
(see Section~\ref{sec-power}); the inner radii for our sample
sources range from 100--450~$R_{\rm G}$, and they are constrained to 
an accuracy of $\approx20\%$.
We note that for J$0401-0540$, the model fits well the red peak
while it shows a slight excess at the blue peak.
In fact, unlike the others, the observed J$0401-0540$ line profile 
possesses a slightly weaker blue peak than
red peak, which cannot be perfectly fit by a circular disc model,
and a model with a non-axisymmetric perturbation is probably required 
\citep[e.g., an elliptical disc; see][]{Eracleous1995}.\footnote{For
the rest of the objects, there may also be some small discrepancies between
the data and model around the peaks. A
perturbed-disc model could likely improve the current fits
by better reproducing the relative strengths of the two peaks.}
However, the inner radius of the emission
region largely depends on the shape of the wings, 
and thus this parameter
determined from a circular disc model is appropriate for deriving the
available local viscous energy.  

\begin{figure}
\centerline{
\includegraphics[scale=0.5]{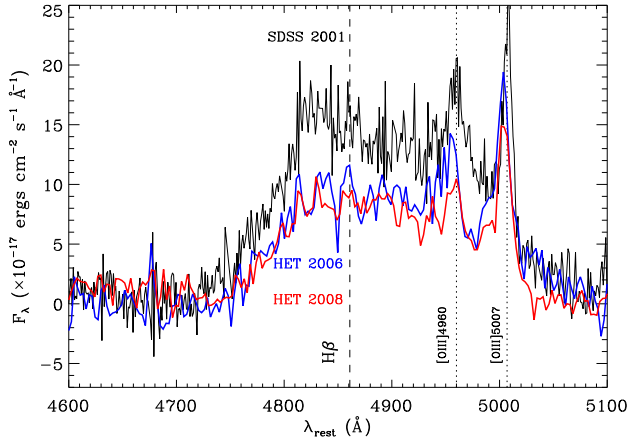}
}
\caption{
Continuum-subtracted H$\beta$ line profiles 
of J$2125-0813$ from the SDSS observation in 2001 (black),
the HET observation in 2006 (blue), and the HET observation in 2008 (red).
The [\ion{O}{iii}] $\lambda5007$ lines appear to have different strengths,
as the spectra have different resolutions and thus the apparent 
widths of the lines
are different; the integrated fluxes are consistent.
The shape of the disc-like H$\beta$ line does not detectably vary. However,
relative to the SDSS observation,
the integrated H$\beta$ line flux dropped by 40\% in 2006 
and 44\% in 2008.
(A color version of this figure is
available in the online journal.)
\label{fig-het}}
\end{figure}

Line profile variability is typical of disc-like emitters
(e.g., \citealt*{Gezari2007}; \citealt{Strateva2008}; \citealt*{Lewis2010}).
We have obtained
additional optical spectra for J$2125-0813$ with the
Low-Resolution Spectrograph \citep{Hill1998} on the
Hobby-Eberly Telescope (HET; \citealt{Ramsey1998})
in 2006 Jun and 2008 Jun, contemporaneous of the \chandra\ or {\it XMM-Newton}
observations.
The disc-like H$\beta$ line profiles in these spectra are
similar to that in the SDSS spectrum taken in 2001, and the 
corresponding
disc-model parameters do not change. However,
the integrated H$\beta$ line flux dropped by 40\% in 2006 and 44\% in 2008
compared to the 2001 value. A comparison of the SDSS and HET spectra
of J$2125-0813$ is shown in Figure~\ref{fig-het}.

Disc-like emission profiles are usually observed in the H$\alpha$
and H$\beta$ lines. Nonetheless, some cases of disc-like
\ion{Mg}{ii} $\lambda2799$ lines are also known
\citep[e.g.,][]{Halpern1996, Strateva2003, Eracleous2003,
Eracleous2004, Eracleous2009, Luo2009}. The SDSS spectra do not cover
the \ion{Mg}{ii} line for the majority of the disc-like emitters in
\citet{Strateva2003} due to their low redshifts. However, the SDSS spectra
of all of the quasars studied here do cover the \ion{Mg}{ii} line, and
the line profiles are displayed in Figure~\ref{fig-opspec}. The
\ion{Mg}{ii} $\lambda2799$ line profiles in the present sample are not
identical to those of the H$\beta$ lines, although a few objects do
appear to have shoulders in their \ion{Mg}{ii} profiles reminiscent of
the shoulders of the H$\beta$ profiles (e.g., J$0959+3449$,
J$1609+0824$, and J$2125-0813$). Moreover, the widths of the
\ion{Mg}{ii} lines are often comparable to those of the H$\beta$ lines.  In
contrast, the \ion{Mg}{ii} profiles of lower-luminosity objects
\citep[studied by][]{Halpern1996, Eracleous2003, Eracleous2004,
Eracleous2009} bear a much stronger resemblance to those of the Balmer
lines, including clearly disc-like shapes. There are several
factors that may cause the differences of the Balmer and \ion{Mg}{ii}
profiles: (1) the \ion{Mg}{ii} line
emissivity is different from that of the Balmer lines
\citep[e.g.,][]{Dumont1990b}; (2) radiative transfer may affect the
Balmer and \ion{Mg}{ii} lines differently due to different optical
depths; (3) the \ion{Mg}{ii} line wings are usually heavily
contaminated by the underlying iron emission-line complexes, making
the derived line profiles less certain; (4) \ion{Mg}{ii} and
\ion{Fe}{ii} absorption complexes can contaminate the \ion{Mg}{ii}
profile and distort its shape \citep[studied by, e.g.,][]{Halpern1996}.
Therefore, there is not always a clear relation between the H$\beta$
and \ion{Mg}{ii} line profiles, and it is more difficult to identify
and study \ion{Mg}{ii} disc-like emitters.

\section{X-RAY AND RELATED DATA ANALYSIS} 
\subsection{X-ray Data Analysis} \label{sec-xray1}
All of the eight disc-like emitters have been
observed by \chandra\ with the S3 CCD of 
the Advanced CCD Imaging
Spectrometer (ACIS; \citealt{Garmire2003}), as listed in Table~\ref{tbl-log}.
For the six Cycle 12 targets, the exposure times were chosen to
yield enough counts ($\ga300$) for basic spectral fitting
in each case, under the assumption that their 
X-ray-to-optical power-law slopes ($\alpha_{\rm OX}$) follow the 
$\alpha_{\rm OX}$--$L_{\rm 2500~\AA}$ relation for typical radio-quiet AGNs
presented in \citet{Steffen2006}.
We reduced and analyzed the ACIS-S3 data
using mainly the \chandra\ Interactive Analysis
of Observations (CIAO) tools.\footnote{See
http://cxc.harvard.edu/ciao/ for details on CIAO.}
We used the {\sc chandra\_repro} script to reprocess the
data with the latest calibration.
The background light curve of each observation was inspected
and background flares were removed using the {\sc deflare}
CIAO script, which utilizes an iterative 3$\sigma$ clipping algorithm.
The background-flare cleaned exposure times are given in 
Table~\ref{tbl-log}.

We created images from the cleaned event files using the standard
\asca\ grade set (\asca\ grades 0, 2, 3, 4, and 6) for the 0.3--8.0 keV band.
We then ran {\sc wavdetect} \citep{Freeman2002} 
on the images to search for X-ray sources, 
with a ``$\sqrt{2}$~sequence'' of wavelet scales (i.e.,\ 1, $\sqrt{2}$, 2,
$2\sqrt{2}$, and 4 pixels)
and a false-positive probability threshold of 10$^{-6}$. 
All eight objects are detected, and their X-ray positions
coincide with the SDSS optical positions ($<0.3\arcsec$ offset; 
see Table~\ref{tbl-log}). They all have 
small off-axis angles (Table~\ref{tbl-log}).

We extracted source counts in three standard bands: 0.5--8.0 keV (full band),
0.5--2.0 keV (soft band), and 2.0--8.0 keV (hard band).
We utilized the polygonal source-count extraction regions generated
by the {\sc acis extract} (AE; \citealt{Broos2010}) software that 
approximate the
$\approx90\%$ encircled-energy fraction (EEF) contour of the point
spread function (PSF) at 1.5 keV.
Background counts were extracted from a source-free 
annular region around each source.
The source counts and the associated 1$\sigma$ errors 
\citep[following][]{Gehrels1986} 
are presented in
Table~\ref{tbl-xspec}. For J$0850+4451$, there are no soft-band photons
within the extraction region, and thus we calculated an upper limit
on the source counts for this band 
using the Bayesian approach of \citet*{Kraft1991} 
for the 99\% confidence level.

The X-ray spectrum of each source was extracted using the CIAO
{\sc specextract} script, with a source aperture of 4\arcsec\ 
radius centered on the X-ray position. The background spectrum was
extracted from an annular region with an inner radius 
of 6\arcsec\ and an outer radius of 15\arcsec, 
which is free of X-ray sources in every case. 
We fitted the spectrum of each source 
using XSPEC
(version 12.6.0; \citealt{Arnaud1996}) except for J$0850+4451$, which has
only three counts in the source spectrum.
We employed an absorbed power-law model ({\sc wabs*zpow*zwabs}) which 
takes into account the Galactic absorption (Galactic neutral
hydrogen column densities are 
listed in Table~\ref{tbl-xspec}; \citealt{Dickey1990})
and possible intrinsic absorption at the redshift of the source. 
If intrinsic
absorption was not required to produce an acceptable fit, 
we estimated the 90\% confidence upper limit
on the intrinsic column density and adopted a simple power-law
model ({\sc wabs*zpow}). In general, the power-law photon index and intrinsic 
column density were set as free parameters during the fit, and
the $\chi^2$ statistic was used. However, 
due to the small number of source counts for J$2148+0010$, 
we fixed its photon index at $\Gamma=1.8$, a typical value for 
luminous 
radio-quiet AGNs \citep[e.g.,][]{Reeves1997,Just2007,Scott2011}, and we also 
adopted the $C$ statistic (cstat) in XSPEC.
The fitting results and fit qualities are presented in Table~\ref{tbl-xspec}.
The quoted errors are at the 90\% confidence level 
for one parameter of interest. For the six sources modeled with the
$\chi^2$ statistic, the best-fitting results
are statistically acceptable,
with the null-hypothesis probability $>27\%$ in all cases.

The X-ray photon indices 
determined from spectral fitting (for six objects) 
have a range
of $\approx1.4$--2.0 (mean $\Gamma=1.75$ with a standard deviation
of 0.28),
typical for luminous radio-quiet quasars.
Considering the established relation between hard \hbox{X-ray} photon
index and H$\beta$ FWHM \citep*[e.g.,][]{Brandt1997,Reeves2000},
we might expect a mean $\Gamma\approx1.7$ for our objects 
(they have a mean H$\beta$ FWHM of $\approx9000$~km~s$^{-1}$).
Our objects thus appear plausibly consistent with this relation,
although our small sample size and the apparently significant intrinsic
dispersion preclude any strict comparison.
Five of the eight sources show no evidence for obscuration.
J$1500+1210$ shows a small amount of absorption, although with a large 
uncertainty, and
J$2148+0010$ is moderately obscured [$N_{\rm H,int}=(2.5_{-1.0}^{+1.2})\times10^{22}$~cm$^{-2}$].
If we adopt a simple power-law model for J$2148+0010$ and
allow its photon index to be a free parameter, 
the resulting value is an unusually flat $\Gamma\approx 0.7$,
also indicating likely obscuration. 
We note that a neutral intrinsic 
absorber was assumed in this fitting. The absorption
column density derived here 
would be an underestimate if the absorbing medium is partially ionized,
but we generally lack sufficient counts to constrain more complex 
absorption models.

We investigated short-term X-ray variability for each source by applying
the Kolmogorov--Smirnov (K--S) test to 
the unbinned event data in the source region
\citep[e.g.,][]{Primini2011}.
The arrival times of events were tested against a constant count-rate model,
and corrections have been made for good time intervals.
None of the sources appears to be variable within the observation, with
the K--S test probability $P_{\rm ks}>0.3$ in every case.
Two of the eight objects, J$0401-0540$ and J$2125-0813$, 
have prior {\it ROSAT} All Sky Survey (RASS)
detections in 1990, allowing
us to examine their long-term X-ray variability. We converted
the RASS 0.1--2.4 keV count rates to the \chandra\ \hbox{0.5--2.0}~keV
count rates using the Portable, Interactive, Multi-Mission Simulator
(PIMMS);\footnote{http://cxc.harvard.edu/toolkit/pimms.jsp.}
photon indices derived from the \chandra\ spectral fitting were adopted.
The RASS-derived count rate (0.078~s$^{-1}$) of J$0401-0540$ appears to be 
consistent with the \chandra\ observation (0.072~s$^{-1}$).
For J$2125-0813$, the count rate observed by \chandra\ (0.056~s$^{-1}$) 
dropped by
a factor of about two over a $\approx20$-year time-scale
compared to that expected from the RASS count rate 
(0.122~s$^{-1}$). Additional \chandra\ and {\it XMM-Newton} 
observations of J$2125-0813$ also indicate long-term variability, as presented
in Section 4.3 below.

\subsection{The Broad Absorption Line Quasar J0850+4451} \label{sec-xray2}

J$0850+4451$ has only three counts within 
the source aperture, preventing spectral modeling. However, these
three counts are all hard X-ray photons, in the \hbox{4--5~keV} range
(corresponding to \hbox{6.2--7.7} keV in the rest frame), suggesting
that this source is heavily obscured. To obtain a first-order estimate of 
the level of obscuration, we 
used the {\sc fakeit} routine in XSPEC to simulate observed spectra
assuming an absorbed power-law model with $\Gamma=1.8$ and
different $N_{\rm H,int}$ values.
A value of 
$N_{\rm H,int}\approx7\times10^{23}$~cm$^{-2}$ is required to generate
three photons with energies above 4 keV and almost no other lower-energy 
photons.
Given the optical luminosity ($L_{\rm 2500~\AA}$) of J$0850+4451$ 
and the $\alpha_{\rm OX}$--$L_{\rm 2500~\AA}$ relation for
typical AGNs (see Section 5.2 for details), we expected $\approx300$ 
full-band counts from this source. The observed number of counts is 
$\approx100$ times smaller than expected, also suggesting heavy obscuration.

\begin{figure*}
\centerline{
\includegraphics[scale=0.5]{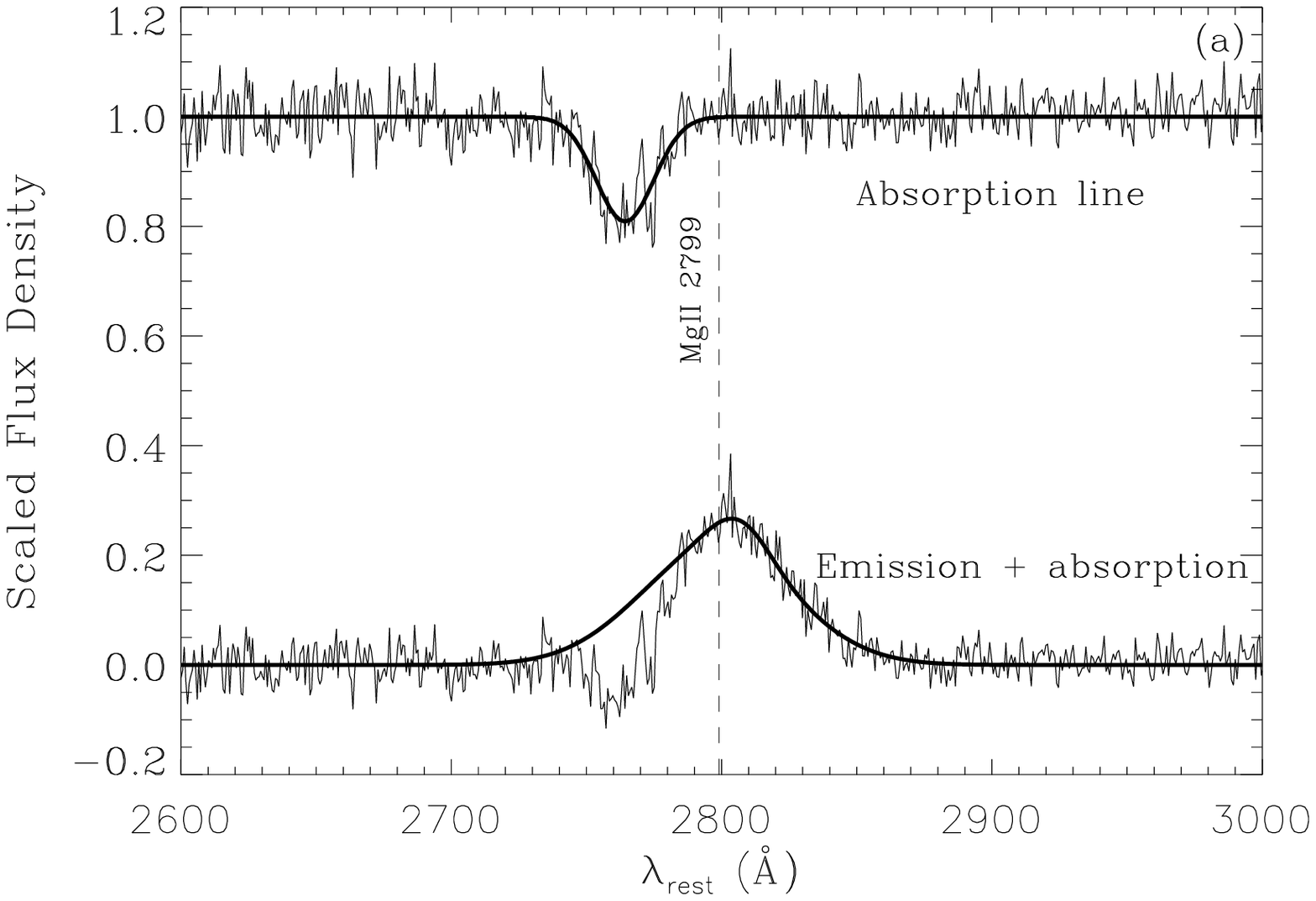}
\includegraphics[scale=0.5]{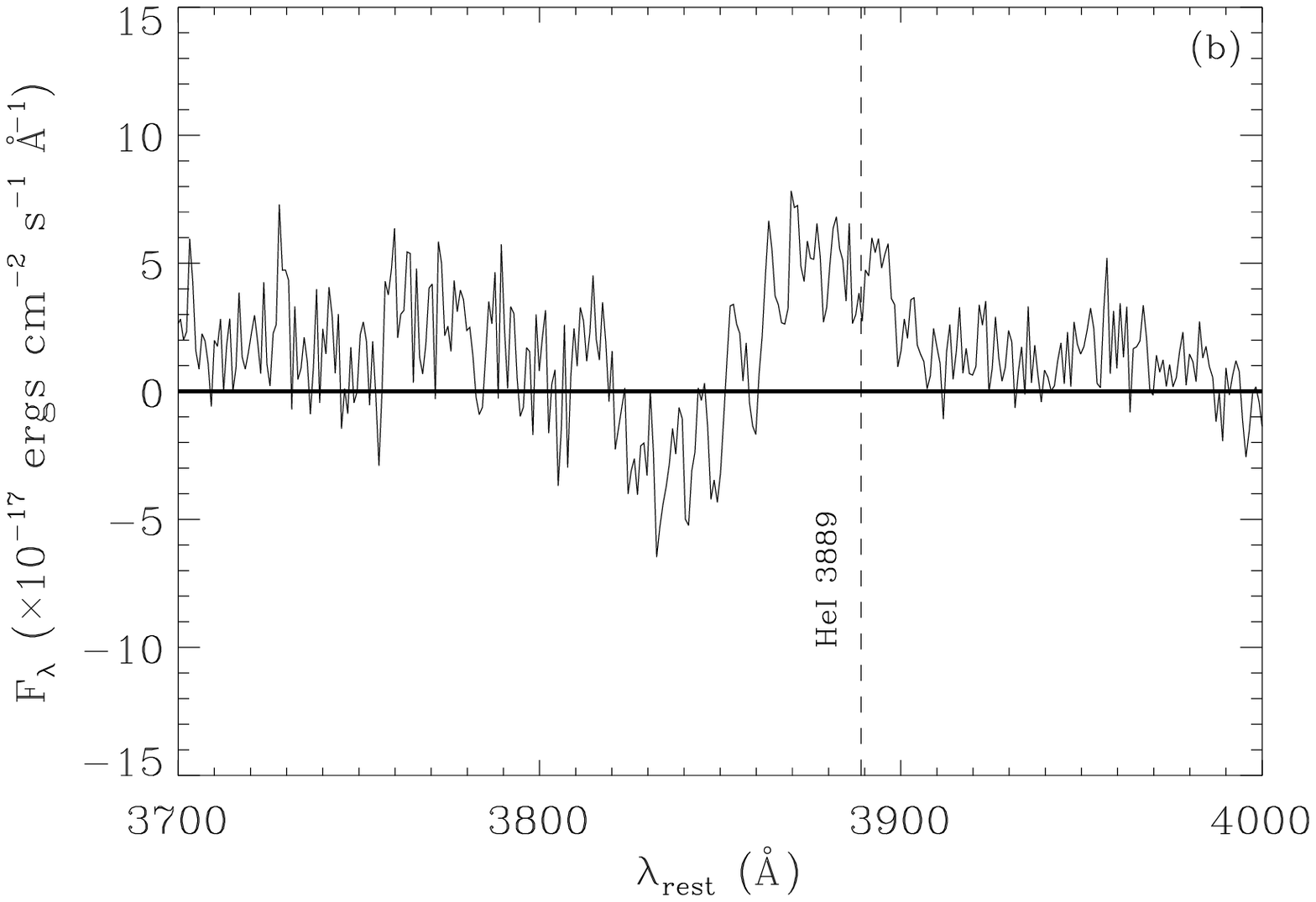}
}
\caption{
Continuum-subtracted (see Section~\ref{sample}) spectra 
around the (a) \ion{Mg}{ii} $\lambda2799$ and (b) \ion{He}{i} $\lambda3889$ 
lines for J$0850+4451$. In (a), the top spectrum shows the absorption line modeled with
one Gaussian profile, after removing the emission line; the bottom
spectrum shows both the absorption and emission along with the 
emission-line model (two Gaussian profiles). The normalizations of the spectra
have been scaled arbitrarily.
\label{fig-bal}}
\end{figure*}

The SDSS spectrum of J$0850+4451$ shows a broad absorption line on the blue 
side of
\ion{Mg}{ii}~$\lambda2799$ with a minimum outflow velocity of 
$v_{\rm min}\approx1300$~km~s$^{-1}$ and a 
velocity width of 
${\rm FWHM}\approx2600$~km~s$^{-1}$, indicating that it is a low-ionization
broad absorption line (LoBAL) quasar 
\citep[e.g.,][]{Boroson1992,Trump2006,Gibson2009}. 
The \ion{Mg}{ii} $\lambda2799$ line is shown in Figure~\ref{fig-bal}a;
We fit two Gaussian profiles to
the broad emission line in the continuum-subtracted spectrum,
and then used one Gaussian to model
the remaining absorption feature.
The SDSS spectrum also shows probable broad
absorption features in \ion{He}{i} $\lambda3188$ and 
\ion{He}{i} $\lambda3889$, 
as have been observed in some other BAL quasars 
(e.g., \citealt{Boksenberg1977}; \citealt*{Leighly2011}).
The \ion{He}{i} $\lambda3889$ line is shown in Figure~\ref{fig-bal}b.
BAL quasars are known to be X-ray weak compared with typical quasars,
while LoBAL quasars are significantly X-ray weaker than even normal BAL quasars,
of which the primary cause is probably strong intrinsic absorption 
\citep[e.g.,][]{Green1995,Green2001,Gallagher2002,Gallagher2006,Gibson2009}.
The heavy X-ray obscuration suggested by the \chandra\ observation is 
consistent with the LoBAL nature of J$0850+4451$. It is the first 
disc-like emitter reported that is also a BAL 
quasar,\footnote{Although UV absorption lines have been observed 
in Arp 102B \citep*[e.g.,][]{Eracleous2003a}, they are relatively narrow
with velocity width $<600$~km~s$^{-1}$.} and thus
disc-like emission lines and BALs are not exclusive phenomena.

\subsection{X-ray Spectra of J2125$-$0813} \label{sec-xray3}

In addition to the \chandra\ observation, we also analyzed the {\it XMM-Newton} observation of J$2125-0813$ (observation
ID: 0505410101) which was taken on 2008 Apr 24. 
The total exposure was 36.3 ks for the MOS
detector and 37.6 ks for the pn detector. We processed the data using 
the standard {\it XMM-Newton} Science Analysis System (v11.0.0) routines.
We reprocessed the raw data (ODFs) to generate calibrated 
and concatenated EPIC event lists, and then filtered these event lists 
for flaring background (${\rm count~rate} >0.5$~s$^{-1}$ for MOS 
and ${\rm count~rate} >0.3$~s$^{-1}$ for pn). The cleaned exposure times
are 20.1 ks (MOS1), 21.0 ks (MOS2), and 22.7 ks (pn).
For each detector, the source spectrum was extracted from a 36\arcsec-radius circular
aperture
centered on the source,
and the background spectrum from a 60\arcsec-radius 
aperture in a nearby source-free region.
The spectra from MOS and pn have $\approx9\,000$ counts in total, and
we jointly fit them using XSPEC assuming 
an absorbed power-law model. 
The resulting best-fitting model is statistically acceptable
($\chi^2/$dof=1.1
and null hypothesis probability $=0.14$), 
and it is
similar 
to that derived from the \chandra\ spectrum, with no intrinsic
absorption ($N_{\rm H,int}<5\times10^{20}~{\rm cm}^{-2}$) and 
$\Gamma=1.45\pm0.05$. The observed 2.0--8.0 keV flux from {\it XMM-Newton}
is $\approx25\%$ smaller than that from \chandra; this type of long-term X-ray 
flux variability
is common among disc-like 
emitters \citep[e.g.,][]{Strateva2006,Strateva2008} and luminous AGNs in
general \citep[e.g.,][]{Gibson2012}.

\begin{figure*}
\centerline{
\includegraphics[scale=0.5]{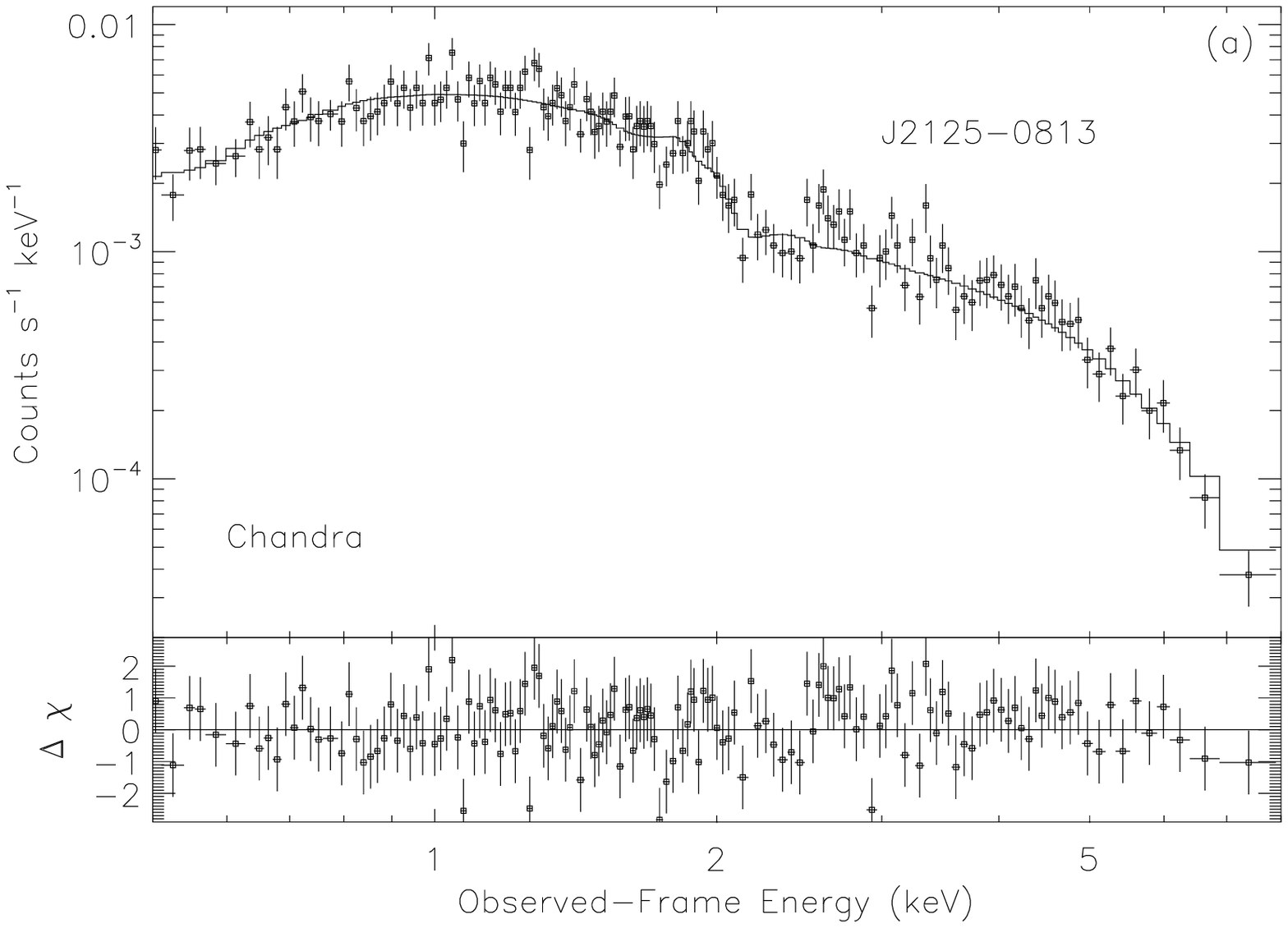}
\includegraphics[scale=0.5]{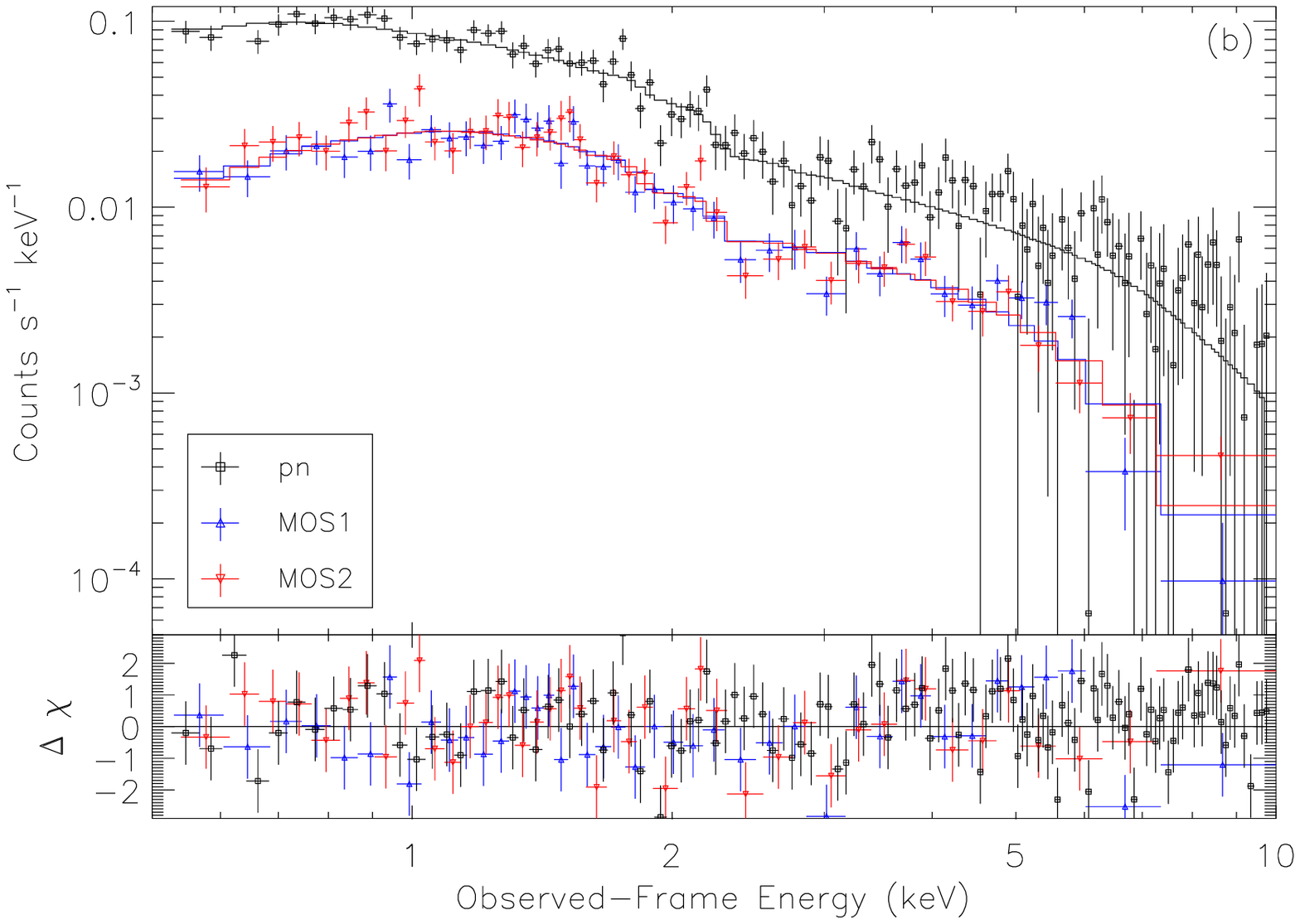}
}
\caption{
X-ray spectra of J$2125-0813$ from (a) \chandra\ and (b) {\it XMM-Newton}
overlaid with the best-fitting models.
The bottom panels show the deviation of the data from the models in units of
$\sigma$ with error bars of size unity. The spectra are modeled
with simple power-law models modified by Galactic absorption; see 
Section~\ref{sec-xray3} for details. There is no apparent 
\ion{Fe}{xxv} K$\alpha$ emission at 6.7 keV
(4.1 keV in the observed frame) as suggested by an earlier 
4~ks \chandra\ exposure \citep{Strateva2008}. 
(A color version of this figure is
available in the online journal.)
\label{fig-xspec}}
\end{figure*}

The X-ray spectra of J$2125-0813$ from \chandra\ and
{\it XMM-Newton}, along with the best-fitting models, are displayed in Figure~\ref{fig-xspec}. Based on an earlier 4~ks \chandra\ observation of this source in 2006 Apr,
\citet{Strateva2008} suggested that there appeared to be a strong
\ion{Fe}{xxv} K$\alpha$ emission line at 6.7~keV with a width of 
0.3 keV and an equivalent width of 700~eV. 
However, the spectrum had only 155 counts and the possible Fe emission
was not reliably constrained.
Given the improved 
\chandra\ and {\it XMM-Newton} observations,
we do not see
any significant excess at 6.7 or 6.4 keV
(4.1 or 3.9 keV in the observed frame) in the observed spectra. 
Compared to the \chandra\ spectrum in \citet{Strateva2008}, the 
\chandra\ X-ray flux
in our analysis 
dropped by $\approx20\%$, within the expected range for typical AGN variability.
We refit the
spectra by adding a Gaussian component at 6.7 (or 6.4) keV 
with a width fixed at 0.3 keV;
the resulting models do not provide improved fits compared to the simple 
power-law models based on $F$-tests. The 90\% confidence upper 
limit on the equivalent width is $\approx300$~eV for a 6.7 (or 6.4) keV 
line.\footnote{The derived upper limit depends on the assumed
line width. In the case of a typical narrow line width of 0.02 keV 
(1000~km~s$^{-1}$), the corresponding upper limit on the equivalent width
is $\approx140$~eV.
In fact, the expected narrow Fe K$\alpha$ equivalent width for a
radio-quiet AGN with a similar X-ray luminosity to J$2125-0813$ is 
$\approx50$~eV, given the X-ray Baldwin effect for Fe K$\alpha$ lines
(e.g., \citealt*{Jiang2006}; \citealt{Wu2009}).}
Therefore, we did not detect
the \ion{Fe}{xxv} K$\alpha$ emission in J$2125-0813$ 
tentatively reported in the previous study.

\section{MULTIWAVELENGTH PROPERTIES}
\subsection{Spectral Energy Distributions} \label{sec-sed}

\begin{figure}
\centerline{
\includegraphics[scale=0.5]{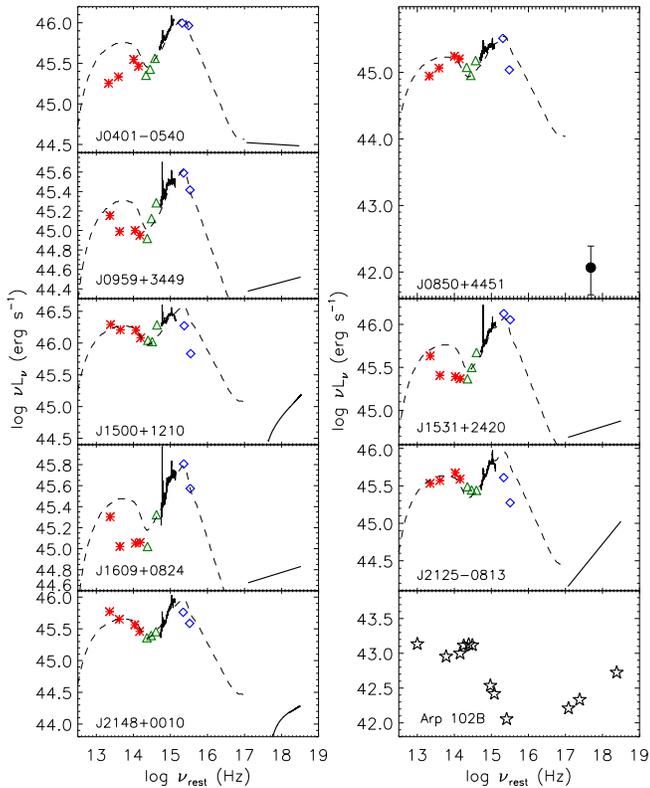}
}
\caption{
IR through X-ray SEDs of the eight disc-like emitters in the rest frame.
IR data (red asterisks) from {\it WISE},
NIR data (green triangles) from 2MASS, the optical spectrum from SDSS, and
UV data (blue squares) from {\it GALEX} are included.
The optical and UV data have
been corrected for Galactic extinction. We adopted the
best-fitting X-ray spectral information in Table~\ref{tbl-xspec}.
The dashed curve
shows the \citet{Richards2006}
radio-quiet quasar
SED template, normalized to the luminosity at rest-frame $10^{15}$ GHz.
The broad-band SED shapes are
in general agreement with the typical radio-quiet quasar SED,
although a few sources show IR deficits (see Section 5.1). The Arp 102B
SED \citep[][and references therein]{Eracleous2003a,Strateva2008}
is also shown for comparison.
(A color version of this figure is
available in the online journal.)
\label{fig-sed}}
\end{figure}

We gathered multiwavelength photometric data for our sample objects to investigate
their spectral energy distributions (SEDs). We obtained infrared (IR) 3.4 $\mu$m,
4.6~$\mu$m, 12.0~$\mu$m, and 22.0~$\mu$m data
from the {\it Wide-field Infrared Survey Explorer} 
({\it WISE}; \citealt{Wright2010}), and
near-infrared (NIR) 
$J$-, $H$-, and $K_s$-band data from the Two Micron All Sky Survey (2MASS; 
\citealt{Skrutskie2006}).
The 2MASS magnitudes in the SDSS DR7 quasar catalog were used; this catalog
provides aperture photometry for additional sources detected down to 2$\sigma$ 
(see Section 5 of \citealt{Schneider2010}).
All eight sources are detected in all the IR--NIR bands except for 
J$1609+0824$ in the $H$ band. We also searched for near-UV (NUV, $\lambda_{\rm eff}=2267$~\AA) and far-UV (FUV, $\lambda_{\rm eff}=1516$~\AA) data from
the {\it Galaxy Evolution Explorer} ({\it GALEX}; \citealt{Martin2005}); all
sources are detected. Using the above data, the SDSS spectra, and 
the \chandra\ best-fitting spectra, we constructed rest-frame 
SEDs for the sample objects,
as shown in Figure~\ref{fig-sed}. The SDSS spectra and UV data 
have been corrected for 
Galactic extinction following the dereddening approach presented in
\citet{Calzetti2000}. 
For J$0850+4451$, we do not have an X-ray spectrum, and 
thus we converted the full-band net counts to the rest-frame 2~keV
luminosity using PIMMS,
assuming a photon index of $\Gamma=1.8$.\footnote{This approach overestimates
the rest-frame 2~keV luminosity as well as the X-ray-to-optical power-law slope 
parameter discussed below (Section~\ref{sec-aox}),
as there were no soft-band photons
detected. However, even the overestimated X-ray luminosity is still
significantly lower than expectations for typical quasars 
(see Section~\ref{sec-aox}).}
We note that the multi-band observations
are not simultaneous, and thus the SEDs are subject to uncertainties caused by
potential source variability (e.g., the optical variability
of J$2125-0813$ in Section~\ref{sec-spec}).

We show in Figure~\ref{fig-sed} the radio-quiet quasar
SED template from \citet{Richards2006}, scaled 
to the luminosities at rest-frame $10^{15}$ GHz (3000~\AA) of our SEDs.
From the NIR to the UV, the SEDs are in good agreement with the template
in general. 
In the IR, a few objects (J$0959+3449$, 
J$1531+2420$, and J$1609+0824$) show 
relatively weaker rest-frame 2--3~$\mu$m emission,\footnote{J$0401-0540$
shows weaker emission at rest-frame 8--14~$\mu$m, different from the other
three IR-weak objects. At these wavelengths, the IR SED may 
have some contribution from the star-formation activity in the
host galaxy \citep[e.g.,][]{Desai2007,Shi2007,Rafferty2011}.} 
by a factor of $\approx1.3$--1.5.
The scatter is within the 1$\sigma$ deviation of the template SED in each case
\citep{Richards2006}.
These sources are luminous quasars, and thus host galaxies should have 
negligible contributions to the optical continuum at rest-frame 3000~\AA, where 
the template scaling was performed; however, the relative scaling 
could have been
affected by optical variability.
The IR SED shapes of the three IR-weak sources are similar 
to those of the class I hot-dust-poor (HDP)
AGNs defined in X-ray selected and optically-selected AGN samples
\citep{Hao2010,Hao2011}, although
the fractions of HDP AGNs in those samples are $\la 10\%$. HDP AGNs are likely
associated with smaller covering factors of dust tori
than typical AGNs, which do not obviously appear to be 
physically connected to the disc-like emission lines that
originate from a much smaller scale disc region.
To assess further the possibility of an intrinsic connection
between IR weakness and disc-like emission lines,
we constructed optical-to-IR SEDs for 47 disc-like emitters
in the \citet{Strateva2003} sample that have a bolometric luminosity
$L_{\rm bol}>10^{45}$~\lum; relatively luminous sources were chosen so that most of them 
would show typical quasar emission from a standard accretion disc.
Among these 47 sources, only three show clear signs of a deficit of
IR emission. Since the host galaxy might have a significant 
contribution to the rest-frame $\approx2$~$\mu$m emission 
when $L_{\rm bol}<10^{45.5}$~\lum\ \citep[e.g.,][]{Stern2012}, we also 
examined the HDP AGN fraction among higher luminosity samples in 
\citet{Strateva2003}. Among the 17 disc-like emitters 
with $L_{\rm bol}>10^{45.5}$~\lum, two ($\approx12\%$) are probably HDP AGNs,
and neither of the two $L_{\rm bol}>10^{46}$~\lum\ objects show an IR deficit.
Therefore, we cannot confirm that the IR weakness observed
in three of our objects is intrinsically connected to their disc-like 
line profiles.

We estimated the bolometric luminosities of the sample objects
by 
integrating
the scaled SED templates and then adding the corresponding 0.5--10.0 keV X-ray 
luminosities. The X-ray luminosities were derived from the XSPEC
models corrected for Galactic absorption or from PIMMS (for J$0850+4451$).
The X-ray-to-bolometric ratios are small,
being $\approx2$--7\% for the five unobscured sources.
The resulting bolometric luminosities are listed in Table~\ref{tbl-fit};
these luminosities have  
uncertainties caused by non-simultaneous observations of the SED data and
potential source variability.

High-luminosity disc-like emitters appear to have 
different SED shapes
from their low-luminosity counterparts \citep[e.g.,][]{Strateva2008,Luo2009}.
We show in Figure~\ref{fig-sed} the SED of Arp 102B for comparison.
The SEDs of our eight objects
show a big blue bump (BBB) in the UV ($\log\nu\approx15.3$),
while the Arp 102B SED is more than two orders
of magnitude fainter and lacks a BBB.\footnote{The IR bump 
of the Arp 102B SED is probably dominated by the emission 
from the host galaxy. However, the argument for the absence of a 
BBB is based on the shape of the SED in the UV through X-ray bands. 
The weakness of the UV continuum cannot be attributed to reddening based 
on the following two considerations: (a) The reddening toward the 
nucleus, as judged from the relative strengths of 
the broad Balmer and
\ion{Mg}{ii} lines, is not high enough to extinguish the UV continuum
and suppress any UV bump \citep{Halpern1996,Eracleous2003}.
In particular, the H$\alpha$/H$\beta$ ratio is 4.3, which is very 
typical of radio-loud AGNs (see Fig. 7 in \citealt{Eracleous2003}), 
i.e., no reddening is required by the data. 
(b) the absorption column density has been measured through
X-ray observations and found to be $\approx3\times10^{21}$~cm$^{-2}$
\citep{Eracleous2003a}.
Models for the absorbing medium by Eracleous et al. (2003) place it 
at a distance from the center comparable to that of the BLR and 
ascribe a high ionization parameter to it, which suggests that 
it is unlikely to be dusty.}
The difference may be attributed to different emission mechanisms 
in high- and low-luminosity sources: Arp 102B is powered by 
a radiatively inefficient accretion flow
\citep[RIAF; e.g.,][]{Rees1982,Narayan1994,Abramowicz2002,Eracleous2003}, while our high-luminosity
objects likely have standard accretion discs that are radiating with
high efficiency \cite[e.g.,][]{Shakura1973}.

We also obtained radio flux information at 1.4~GHz for these sources.
For sources detected by the Faint Images of the Radio Sky at 
Twenty-Centimeters (FIRST) survey \citep*{Becker1995},
the fluxes were taken from the FIRST source catalog \citep{White1997}. 
For sources not detected but still covered by the FIRST survey, the 
upper limits on the radio fluxes were set to $0.25+3\sigma_{\rm rms}$ mJy,
where $\sigma_{\rm rms}$ is the rms noise of the FIRST survey at the 
source position and 0.25 mJy is to account for the CLEAN bias 
\citep{White1997}. For the one source (J$0401-0540$) that 
is not covered by the FIRST survey, we checked the 
NRAO VLA Sky Survey (NVSS; \citealt{Condon1998}) catalog and determined its flux upper limit to be
1.35~mJy, which is three times the rms noise of the NVSS. 
Utilizing the optical and radio data, We calculated 
the radio-loudness parameters,
defined as $R=f_{5~{\rm GHz}}/f_{\rm 4400~\AA}$ 
\citep[the ratio of
flux densities in the rest frame; e.g.,][]{Kellermann1989}.
The 5~GHz and 4400~\AA\ flux densities were converted from the
observed 1.4 GHz and rest-frame 2500~\AA\ flux densities assuming a radio 
power-law slope of $\alpha_{\rm r}=-0.8$ 
(e.g., \citealt*{Falcke1996}; \citealt{Barvainis2005})
and an optical power-law slope of $\alpha_{\rm o}=-0.5$
($f_\nu\propto \nu^{\alpha}$; e.g., \citealt{Vandenberk2001}). The radio-loudness parameters or the 
upper limits are listed in Table~\ref{tbl-aox}; all eight sources
are radio quiet ($R<10$) with most having $R<2$, in accordance with our
sample selection in Section 2.

\subsection{Distribution of the X-ray-to-Optical Power-law Slopes} \label{sec-aox}

\begin{figure}
\centerline{
\includegraphics[scale=0.5]{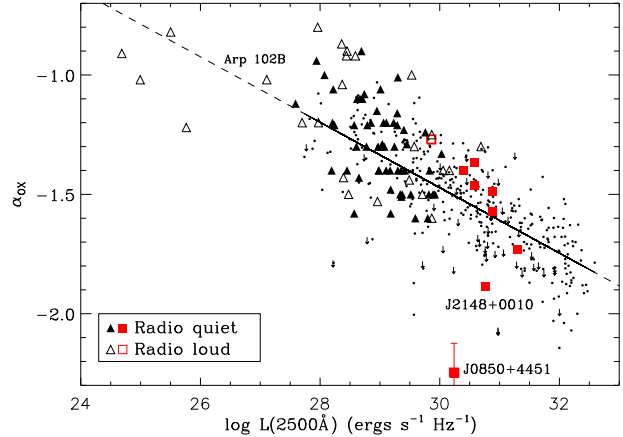}
}
\caption{
X-ray-to-optical power-law slope vs. 2500 \AA\ monochromatic luminosity.
The solid line is
the \citet{Steffen2006} $\alpha_{\rm OX}$--$L_{\rm 2500~\AA}$ relation with
the corresponding sample represented by the dots and downward arrows 
(upper limits).
The dashed line is an extrapolation of the \citet{Steffen2006} relation to
lower luminosities. The open (filled) triangles represent the
radio-loud (radio-quiet) disc-like emitters
from previous studies \citep{Eracleous2003,Strateva2003,Strateva2006,Strateva2008}.
The red filled squares indicate the eight objects in
our sample, and the red open square represents J$0331-2755$. 
The errors on $\alpha_{\rm OX}$ for our disc-like emitters
were propagated from the errors
on the X-ray counts; they are smaller than or comparable to the symbol size
and are not visible except for the case of 
J$0850+4451$. (A color version of this figure is
available in the online journal.)
\label{fig-aox}}
\end{figure}

The X-ray-to-optical power-law slope parameter 
[$\alpha_{\rm OX}=-0.3838\log(f_{2500~{\rm \AA}}/f_{2~{\rm keV}})$]
measures the X-ray brightness of a quasar relative to its optical/UV luminosity.
We calculated the $\alpha_{\rm OX}$ parameters for our sample objects and 
compared them to the expected values derived from
the \citet{Steffen2006} $\alpha_{\rm OX}$--$L_{\rm 2500~\AA}$ relation
for radio-quiet quasars; the $\alpha_{\rm OX}$ values and offsets are
reported in Table~\ref{tbl-aox}. In Figure~\ref{fig-aox}, we plot
the locations of our sources and the \citet{Steffen2006} 
$\alpha_{\rm OX}$--$L_{\rm 2500~\AA}$ relation;\footnote{We 
extrapolated the \citet{Steffen2006} relation to lower luminosities
than those originally used to define it.
However, the $\alpha_{\rm OX}$ values for low-luminosity AGNs 
will probably lie below the extrapolation as suggested by,
e.g., \citet{Steffen2006} and \citet{Maoz2007}.
The extrapolation was for illustrative purposes only and does not affect
our analysis here.} 
also shown are some 
disc-like emitters from previous studies 
\citep{Eracleous2003,Strateva2003,Strateva2006,Strateva2008}.
Our eight sources reside at the high-luminosity end, and their
$\alpha_{\rm OX}$ parameters follow the $\alpha_{\rm OX}$--$L_{\rm 2500~\AA}$ relation (within 1$\sigma$) except for
J$0850+4451$ and J$2148+0010$, which are likely X-ray weak due to 
the previously noted absorption (see Section 4). 
The X-ray luminosity of J$0850+4451$ is $\approx80$ times smaller than
that expected from its optical luminosity.
The highest-redshift disc-like emitter, J$0331-2755$, has 
a typical $\alpha_{\rm OX}$ value despite its radio-loud nature.

For the overall radio-quiet disc-like emitter population,
there is no sign of significantly diminished or enhanced X-ray emission observed
in the $\alpha_{\rm OX}$--$L_{\rm 2500~\AA}$ plot, and thus the 
disc-like line profiles are not connected to any level of excess 
X-ray emission. This conclusion is consistent with the finding in 
\citet{Strateva2006}. While \citet{Strateva2008} suggested that the
disc-like emitters in their sample have excess X-ray emission relative 
to the optical/UV emission, that result was based on only four radio-quiet
objects which are exceptional
disc-like emitters with the very broadest Balmer lines 
(${\rm FWHM}>14\,000$~km~s$^{-1}$). 
Further investigation is needed to assess if the X-ray emission
is unusually strong in the broadest disc-like Balmer line emitters
or if the \citet{Strateva2008}
result is simply due to small-number statistical fluctuations.

\section{DISCUSSION} \label{sec-discuss}
\subsection{Energy Budget Requirements for the Line Emission} \label{sec-power}

The energy budget problem is present in many disc-like emitters,
as their H$\alpha$ line luminosities are close to or even exceed the 
total gravitational power dissipated in the \hbox{line-emitting} 
region of the accretion disc 
(e.g., \citealt*{Chen1989b}; \citealt{Eracleous1994,Strateva2006}).
Following Equation 4 of \citet{Eracleous1994}, which was derived by
integrating Equation 20 of \citet{Chen1989b},
we estimated the viscous
power released in the \hbox{line-emitting} region ($W_{\rm d}$) to be
\begin{equation}
\begin{split}
W_{\rm d}=7.7\times L_{\rm bol}&\left[ \frac{1}{R_{\rm in}}\right.\left(
1-\sqrt{\frac{8}{3R_{\rm in}}}\right) \\
&\left. -\frac{1}{R_{\rm out}}\left(
1-\sqrt{\frac{8}{3R_{\rm out}}}\right)\right]~{\rm erg~s^{-1}}.
\end{split}
\end{equation}
An accretion efficiency
of $\eta\approx0.1$ ($L_{\rm bol}=\eta\dot{M}c^2$) was assumed for this
equation.
The inner and outer radii of the emission region are from
the disc-model fit in Section~\ref{sec-spec}. 
The derived energy budget of the disc region was then compared to
the H$\beta$ luminosity, integrated from the best-fitting model.
The $L_{\rm H{\beta}}$ and $W_{\rm d}$ values 
are listed in Table~\ref{tbl-fit}; the $L_{\rm H{\beta}}$ to $W_{\rm d}$ 
ratios have a range of 
0.02--0.13 for our sample objects, with a median value of 0.05.

\begin{figure*}
\centerline{
\includegraphics[scale=0.5]{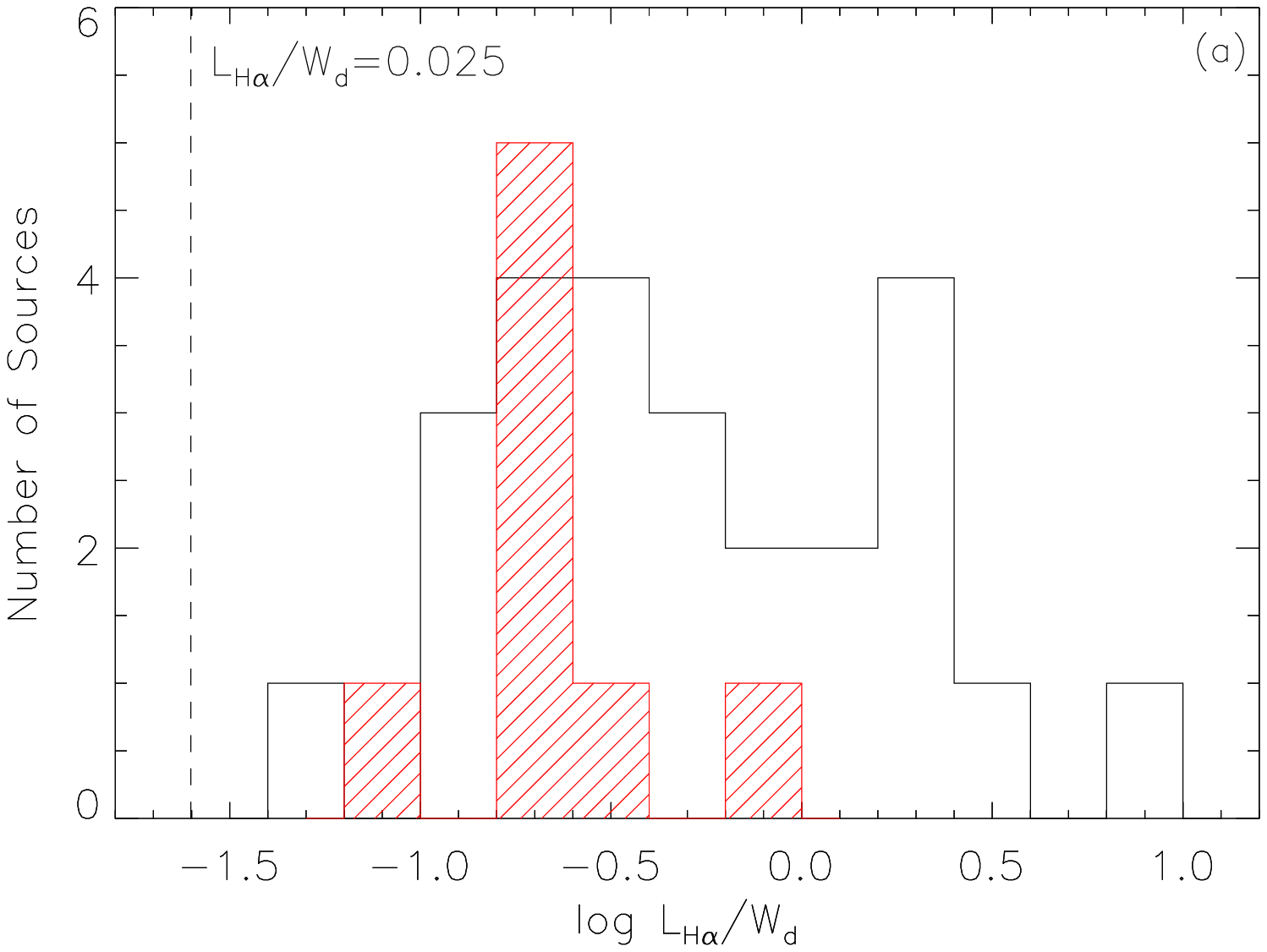}
\includegraphics[scale=0.5]{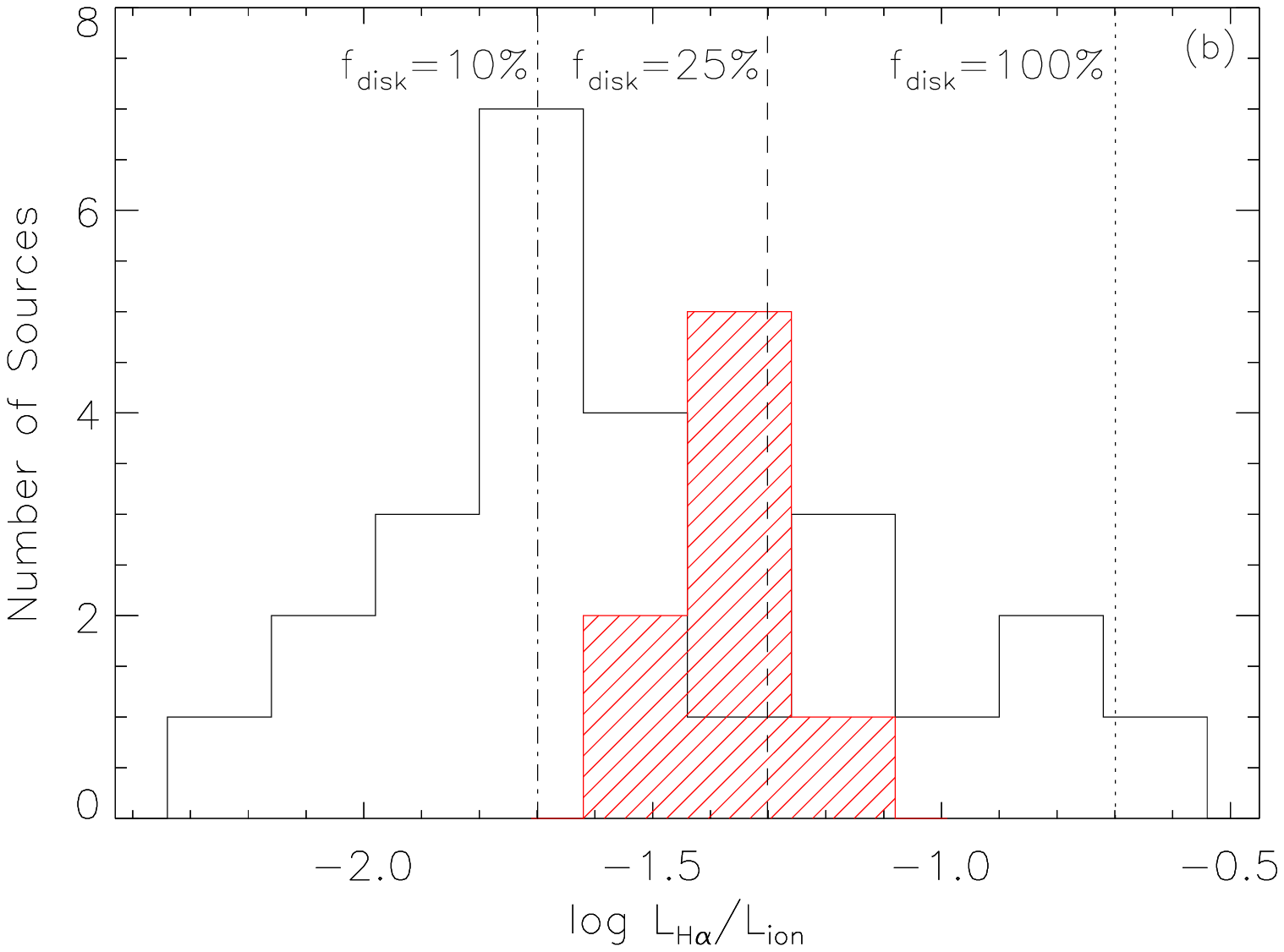}
}
\caption{
Histograms of (a) the ratio of the
H$\alpha$ luminosity to disc gravitational power
and (b) the ratio of the H$\alpha$ luminosity to the
ionizing luminosity ($L_{\rm ion}=0.2L_{\rm bol}$).
The black histograms are for 25
disc-like emitters (including Arp 102B) from \citet{Eracleous2003},
\citet{Strateva2008}, and \citet{Luo2009}. The red filled
histograms are for the eight objects in our sample.
The H$\alpha$ luminosities for our sources
were converted
from the H$\beta$ luminosities as described in Section~\ref{sec-power}.
In (a), the vertical dashed line indicates the upper limit of the 
$L_{\rm H\alpha}$ to $W_{\rm d}$ ratio
if the line emission is solely powered by the gravitational energy
of the disc-emission region.
In (b),
the bolometric luminosities
for the sources in \citet{Eracleous2003}
were derived from their soft X-ray luminosities assuming a bolometric 
correction of 10; for Arp 102B, the bolometric luminosity is from
\citet{Lewis2006}.
The dotted, dashed, and dash-dotted lines indicate the 
expected $L_{\rm H\alpha}/L_{\rm ion}$ values
if 100\%, 25\%, and 10\% of the ionizing luminosity from the 
AGN is utilized to ionize the line-emitting region of the disc
and power the line emission, respectively.
The median required
$f_{\rm disc}$ value is $\approx15\%$, and it is
$\approx22\%$ for our high-luminosity objects.
(A color version of this figure is
available in the online journal.)
\label{fig-enbudget}}
\end{figure*}

Assuming that the line emission is solely powered by the gravitational energy
dissipated locally,\footnote{Here we adopt the \citet{Shakura1973}
notion of local energy dissipation, i.e., that torques in the disc
convert kinetic energy to heat, which is deposited locally in the gas as
thermal energy.} we can derive the predicted ratio of H$\beta$ luminosity 
to total viscous power 
following Equations 3.3 and 3.4 of \citet{Dopita1996}. These authors
calculate the radiative cooling via Balmer line emission
of thermal plasmas for a wide range of conditions and the results
are not dependent upon how
the plasmas were heated.\footnote{The thermal plasma assumption
was adopted here only to assess if the disc-like line emission 
can be solely powered by the local gravitational energy.}
For a typical disc temperature of $T\approx10^4$--$10^5$~K, 
the expected ratio of H$\beta$ luminosity to 
total power is about 0.003--0.005.\footnote{The expected $L_{\rm H\beta}$ to
$W_{\rm d}$ ratio is estimated to be $\la0.06$ in \citet{Strateva2008}.
However, that ratio was derived from \citet{Williams1980}, who presents
only the relative ratios of emission lines (i.e., continuum emission is 
not included). Therefore, 0.06 is an
extremely conservative upper limit on the $L_{\rm H\beta}$ to
$W_{\rm d}$ ratio.} 
The ratio would be even lower if there are radiative-transfer 
effects in the emission region \citep[e.g., see Fig.~5 of][]{Murray1997}.
Therefore, the expected 
ratio of H$\beta$ luminosity to
total power is much smaller than the observed values above (\hbox{0.02--0.13}), indicating
that our sample objects require additional energy 
(likely in the form of external illumination) to power the 
line emission. 
In Figure~\ref{fig-enbudget}a, we show the 
distribution of 
$L_{\rm H{\alpha}}$ to $W_{\rm d}$ ratios of our eight objects and 
25 disc-like emitters (including Arp 102B) from \citet{Eracleous2003},
\citet{Strateva2008}, and \citet{Luo2009}, where the 
H$\alpha$ luminosities for our sources
were converted 
from their H$\beta$ luminosities assuming a typical Balmer decrement of 5 
($L_{\rm H{\alpha}}/L_{\rm H{\beta}}=5$).\footnote{
This Balmer decrement was derived by taking the
average of $L_{\rm H{\alpha}}/L_{\rm H{\beta}}$ values for luminous
disc-like emitters in \citet{Eracleous2003}. The standard deviation
of this average is about 1.} 
Given such a Balmer decrement, the ratio of H$\alpha$ luminosity to
disc viscous energy should not exceed 0.025 for a thermal plasma, and thus
all these disc-like emitters
require external illumination of the accretion disc.

It is therefore likely that the accretion disc is 
photoionized and a thin ``skin'' emits the observed 
lines. The source of photoionization
is considered to be the UV-to-X-ray emission produced in the inner
disc/corona \citep[e.g.,][]{Chen1989}. We calculated
the available ionizing luminosity ($L_{\rm ion}$) 
by integrating the source SED blueward of 
13.6~eV. For our luminous disc-like emitters, the SED shapes are 
similar, and the ionizing-to-bolometric luminosity ratios are $\approx0.2$.
As the X-ray-to-bolometric ratios are small (see Section~\ref{sec-sed}),
the ionizing luminosity is dominated by extreme UV 
emission.\footnote{The majority of the extreme UV portion of 
the \citet{Richards2006}
template SED was, by necessity, not covered
by observational data, and was instead derived by interpolating 
the UV and soft X-ray data. The estimated 
ionizing luminosity is therefore subject to some uncertainty.}
For low-luminosity sources, such as Arp 102B, the SED shapes 
are different (see Figure~\ref{fig-sed}); however,
the ionizing-to-bolometric luminosity ratios are also coincidentally
$\approx0.2$, and
the ionizing luminosity is dominated by X-ray emission.
In Figure~\ref{fig-enbudget}b, we show the distribution of 
$L_{\rm H{\alpha}}$ to $L_{\rm ion}$ ratios for our eight objects and
the 25 disc-like emitters from literature (the same sample as used
in Fig.~\ref{fig-enbudget}a).
The H${\alpha}$ line luminosity produced by the photoionized disc
depends on the fraction of the ionizing energy that is intercepted by
the line-emitting region of the disc (we define this fraction as $f_{\rm disc}$) 
and the fraction of the input energy that can be emitted as H$\alpha$.
For the latter, the fraction is estimated to be
$\approx0.2$ for a photoionized disc 
(Section 4.2 of \citealt{Collin1989}). Using the above relation, 
we plot in Figure~\ref{fig-enbudget}b the 
expected $L_{\rm H\alpha}/L_{\rm ion}$ values when $f_{\rm disc}=100\%$,
25\%, or 10\%. It appears that
for a significant fraction of the disc-like emitters, $f_{\rm disc}>10\%$
is required for the UV-to-X-ray emission to ionize the outer disc and produce
the Balmer lines. In some cases, $f_{\rm disc}=25\textrm{--}100\%$ 
is probably required. The median required 
$f_{\rm disc}$ value is $\approx15\%$
for all the sources shown in Figure~\ref{fig-enbudget}b, and it is 
$\approx22\%$ for our sample objects.
We caution that due to various uncertainties in the estimations of the 
luminosities and luminosity relations, it is not feasible to
calculate the precise $f_{\rm disc}$ values for individual sources, and the 
requirement above is only valid in a statistical sense.

For the general population of AGNs, similar energy budget 
analyses considering the line luminosity
and available ionizing power have also placed constraints upon
the covering factor of the BLR, which is estimated to be
$\approx10$--40\% (see, e.g., \citealt{Maiolino2001} and 
references therein). 
In particular, \citet*{Korista1998} have been able to reproduce 
the equivalent widths of the UV emission lines and 
the Baldwin effect in the context of the "locally optimally emitting cloud"
model with a covering factor of 0.5.
Under the traditional picture of the BLR, where
it consists of discrete ``clouds'' surrounding the nucleus and it is 
illuminated directly by the ionizing continuum,
such a covering factor can be straightforwardly achieved. Additionally, the 
lack of Lyman continuum absorption (e.g., \citealt{MacAlpine2003}
and references therein) and the high covering factor imply that 
the BLR does not cover the nucleus uniformly, and it has been
suggested that the BLR is flattened in the plane of the disc
\citep[e.g.,][and references therein]{Maiolino2001,Gaskell2009}.
For disc-like emitters, the BLR (at least for the low-ionization broad lines) 
is thought to be the skin of the outer accretion disc, and 
its structure has much less uncertainty/flexibility. The radius of the BLR
ranges from a few hundred to a few thousand $R_{\rm G}$, and the
vertical extent is very limited for a standard disc.
Therefore, the requirement upon $f_{\rm disc}$ places strong geometric 
constraints on the inner structure of the AGN central engine.

The emission from the inner disc can ionize the line-emitting 
region of the outer disc (BLR) via direct illumination.
For low-luminosity disc-like emitters (e.g., Arp 102B), 
the inner disc may have a vertically extended structure, 
such as a geometrically thick and
optically thin RIAF.
\citet{Chen1989} calculated that the outer disc can subtend 
a maximum solid-angle fraction of 21\%, assuming that the inner disc
is a spherical, optically thin illuminating source with radius 
$r_0=R_{\rm in}$. For the high-luminosity sources in our sample, 
the accretion disc is 
likely geometrically thin (see Section~\ref{sec-sed} for SED comparisons
of high- and low-luminosity sources), and thus the outer disc
should subtend a much smaller solid angle provided the disc is flat. 
It appears that direct illumination for a flat disc cannot provide
enough energy to power the observed Balmer lines in some 
disc-like emitters, both high-luminosity and low-luminosity, 
as indicated by the required $f_{\rm disc}$ values shown in 
Figure~\ref{fig-enbudget}b. A warped accretion disc
induced by nuclear radiation 
(e.g., \citealt*{Maloney1996}; \citealt{Pringle1996})
or other effects
may increase the solid angle subtended by the line-emitting region of
the outer disc and provide a possible solution for the energy budget problem 
\citep*[e.g.,][]{Wu2008}.

Another possible mechanism for redirecting ionizing photons toward
the outer disc (BLR) is scattering,
where the emission from the inner disc is scattered back 
toward the outer disc by an ionized medium
\citep[e.g.,][]{Dumont1990}.
For low-luminosity disc-like emitters, 
the ionizing luminosity is dominated by X-ray emission, and
the scattering gas may be from a jet
or outflow \citep{Cao2006}. 
For the high-luminosity sources presented here, 
the ionizing luminosity is dominated by UV emission, and
the scattering medium must be highly ionized 
in order to scatter UV photons efficiently.\footnote{The small radio-loudness 
parameters of our eight objects indicate that radio jets 
play a minor role, if any, in the scattering process.}
In either case, significant
X-ray obscuration, probably highly ionized, would be seen if 
the observer's line of sight passes 
through the scattering
gas. Among the limited samples studied with \chandra\ or 
{\it XMM-Newton} in \citet{Strateva2006}, \citet{Strateva2008}, and this paper,
the fraction of disc-like emitters with significant intrinsic absorption 
($N_{\rm H,int}>10^{22}$~cm$^{-2}$)
is $\approx20$--40\%, which constrains the covering factor of the 
scattering gas.\footnote{Only type 1 AGNs are considered here.
Therefore, the covering factor might also depend on the ratio of 
type~1 to type~2 AGNs. Assuming a type 1 to type 2 ratio of $\approx1$ 
for high-luminosity AGNs \citep[e.g.,][]{Reyes2008,Lawrence2010} 
and that all type 2 AGNs have
the scattering medium, the covering factor could be as high as 
$\approx70\%$.} 
A neutral intrinsic absorber was generally assumed to estimate
the column densities for these sources, and the column densities would
be higher if the medium is highly ionized.
The fraction of the total 
ionizing luminosity reaching the outer disc depends on
the gas covering factor, 
the Thomson optical depth of the gas, and the scattering geometry. 
Assuming a covering factor of $40\%$, an optical depth of $\tau_{\rm T}=0.2$
(corresponding to $N_{\rm H}\approx3\times10^{23}$~cm$^{-2}$), 
and a geometric factor
of $<0.5$ (0.5 seems the absolute maximum possible value given the optically thin nature
of the scattering gas),
the $f_{\rm disc}$ value provided by scattering is $<4\%$.
Given the $f_{\rm disc}$ requirement shown in
Figure~\ref{fig-enbudget}b, scattering alone is probably not sufficient
to provide the required illumination of the line-emitting disc.
A systematic \chandra\ or {\it XMM-Newton} spectral 
survey of a sample of
disc-like emitters, obtaining sufficient counts ($\approx10\,000$) 
per object,
could provide further constraints on 
the optical depth and covering factor of any putative scattering
medium.

The
gravitational energy dissipated locally may also collisionally ionize
some fraction of the atoms in the disc skin and contribute to solving
the energy budget problem \citep[e.g.,][]{Collin1987}. 
Given the expected $L_{\rm H{\beta}}$
to $W_{\rm d}$ ratio above under the assumption of an emission region
solely powered by viscous energy, we estimate that $\la10\%$ of the
observed H$\beta$ luminosity
could be produced by this type of collisional ionization for our objects
here. Therefore, the emission lines appear to be mainly powered 
by direct illumination and scattering processes. 
On average, $\approx15\%$ ($\approx22\%$ for our high-luminosity objects here) 
of the nuclear ionizing radiation is utilized to photoionize the line-emitting
disc, although the mechanism is still
not clear.
Warped accretion discs are probably needed for direct illumination to
work efficiently, especially in high-luminosity objects.

\subsection{Connection with the General Population of AGNs}

Disc-like emitters are closely related to the more general
population of AGNs with single-peaked broad lines. 
The general properties of disc-like emitters are similar to those
of typical AGNs, except for the disc-like and generally broader
emission lines.
In fact, a few AGNs have shown broad Balmer lines that fluctuate between
a disc-like and a single-peaked profile on time-scales of years, 
e.g., NGC 5548 
\citep[e.g.,][]{Peterson1999,Sergeev2007},
Pictor A \citep[e.g.,][]{Halpern1994,Sulentic1995}, and Akn 120 \citep*[e.g.,][]{Alloin1988}.
The eight objects presented here sample the high-luminosity and
relatively high-redshift parameter space, 
and they still possess typical AGN SEDs and X-ray spectra.
Since all relatively luminous AGNs are expected to have accretion
discs, and such discs are capable of producing 
disc-like emission lines, it is puzzling why the majority of
AGNs show single-peaked lines. 
The parameters of the line-emitting region, such as the 
disc inclination angle and the ratio of the inner to outer radius,
do in principle affect the 
appearance of the line profile and in some cases result in single-peaked 
lines, though these alone may not be
sufficient to
explain the rare occurrence of disc-like emitters (see Section 1).

One promising model is the disc-wind scenario \citep[e.g.,][]{Murray1997,Eracleous2003,Chajet2012,Flohic2012}, 
where radiative-transfer
effects through a disc wind in and above the line-emitting region of the disc
can alter the emergent profile and produce a single-peaked
line. Radiatively driven disc winds have long been 
suggested to be a common feature of AGNs \citep[e.g.,][]{Murray1995}.
Calculations by \citet{Flohic2012} and \citet{Chajet2012} 
show that the appearance of the
observed line depends mainly on the optical depth of the wind
and the line-emitting region parameters;
a large optical depth ($\tau\ga10$) could result in a single-peaked line profile.
With this model, the transitions from a single-peaked line to a
disc-like line
observed in a few sources can
be naturally explained by a change in the optical depth of the wind.
The observed small fraction ($\approx3\%$) of disc-like emitters among
AGNs suggests that most AGNs have winds with relatively large optical depths.
Moreover, since the wind is radiatively driven,
AGNs radiating at close to the Eddington limit (likely
high-luminosity AGNs)
should generally have stronger winds with larger optical depths
than AGNs with low Eddington ratios (low-luminosity sources; e.g., \citealt{Ganguly2007}).
Therefore, the fraction of disc-like emitters should decline with
AGN luminosity. 
We found 14 disc-like emitters within the 3132 luminous AGNs searched 
(see Section~\ref{sample}), corresponding to a fraction of 0.4\%, although
our selection may not be complete owing to the fact that we only
searched for secure disc-like emitters and may have missed some
possible candidates (also see Footnote 3). 
Nevertheless, this derived fraction is 
a factor of $\approx7$ smaller than
the $\approx3\%$ for the lower-luminosity sample
in \citet{Strateva2003}, likely consistent with the basic expectation
from the disc-wind model. Moreover, we have not found 
any profiles with widely separated peaks, reminiscent of those 
of low-luminosity objects such as Arp 102B.
To compare the fractions of disc-like emitters among
AGNs in different luminosity bins robustly, they
must be identified in a uniform and systematic 
way for both the high-luminosity
and low-luminosity samples, utilizing quantitative selection criteria.
This is beyond the scope of the current paper.

The discovery of a BAL quasar, J$0850+4451$, in our disc-like emitter
sample is notable, as BAL quasars are also a sub-type of 
AGNs. BALs have been observed in $\approx15\%$ of optically-selected quasars \citep[e.g.,][]{Hewett2003,Trump2006,Gibson2009}. Similar to disc-like
emitters, BAL quasars have indistinguishable IR-to-UV SEDs from non-BAL quasars
\citep[e.g.,][]{Gallagher2007}, although BAL quasars usually show 
X-ray absorption as mentioned in Section~\ref{sec-xray2}. A popular 
interpretation is that BAL quasars and non-BAL quasars 
are the same type of object but with different viewing angles 
\citep[e.g.,][]{Weymann1991,Ogle1999},
or at least that orientation plays an important role in determining the 
presence of BAL features \citep*[e.g.,][]{DiPompeo2012}. It is suggested that 
BALs are observed when our line of sight passes through the disc
wind \citep[e.g.,][]{Murray1995}.
Traditionally,
such models have required that BAL quasars generally be viewed edge-on,
with large inclination angles ($i\approx60\degr$).
However, more recent results 
(\citealt{DiPompeo2012} and references therein) suggest that BAL quasars
can be seen over a wide range of inclination angles, with a tendency
to be seen at larger inclination angles than non-BAL quasars. In the 
disc-wind scenario, this result constrains disc winds to flow out at different
angles above the disc in different objects.
The disc inclination angle of J$0850+4451$ derived by the disc-model fit
in Section~\ref{sec-spec} 
is $34\degr$, the largest value of $i$ in our sample.
Thus, it fits in with the result that BAL quasars tend to be seen at
larger inclination angles than non-BAL quasars.

\section{SUMMARY AND FUTURE WORK}

The main results of this study are the following:

\begin{enumerate}

\item
We measured the X-ray and multiwavelength properties of eight 
high-luminosity
($L_{\rm 2500~\AA}>10^{30}$ \mlum) and relatively
high-redshift ($z\approx0.6$) disc-like emitters selected from 
SDSS DR7. These sources are located in 
a largely unexplored space of the luminosity-redshift plane.
See Section 2.

\item
Our sample objects show disc-like H$\beta$ line profiles that can be 
well explained by emission from the outer region (hundreds
to thousands of $R_{\rm G}$) of a Keplerian accretion disc.
The emission-region 
parameters are within the typical ranges found for other disc-like
emitters. See Section~3.

\item
Seven sources have typical X-ray spectra with photon indices 
$\Gamma\approx1.4$--2.0; two of them show some amount of X-ray absorption
($N_{\rm H}\approx10^{21}$--$10^{22}$~cm$^{-2}$ when modeled using neutral gas).
The other source, J$0850+4451$,
has only three $>4$~keV photons observed and is probably heavily obscured
($N_{\rm H}\ga3\times10^{23}$~cm$^{-2}$). Based on the longer observations of 
J$2125-0813$ by \chandra\ and {\it XMM-Newton},
we did not detect the tentative
\ion{Fe}{xxv} K$\alpha$ line emission reported by \citet{Strateva2008}.
See Section~4.

\item
We constructed IR-to-UV SEDs of the sample objects. All
have typical quasar SEDs with a BBB in the UV, different from
low-luminosity disc-like emitters. The emission mechanisms
are probably different: low-luminosity sources are powered by RIAFs, while
high-luminosity sources have standard accretion discs radiating with
high efficiency. See Section~5.1.

\item
The X-ray-to-optical power-law slope parameters ($\alpha_{\rm OX}$) for our 
sources follow 
the established \hbox{$\alpha_{\rm OX}$--$L_{\rm 2500~\AA}$} relation except for 
the two sources with significant X-ray absorption. 
There does not appear to be any significant excess X-ray emission for the 
radio-quiet disc-like emitter population overall.
This high-luminosity and relatively
high-redshift sample does not show any distinct features 
in terms of X-ray emission or multiwavelength properties.
See Section 5.2.

\item
We analyzed the energy budget requirements for powering the emission lines.
For disc-like emitters in general, 
external illumination is required.
The inner disc
should illuminate and ionize the outer disc efficiently via
direct illumination and/or scattering;
the median required
$f_{\rm disc}$ value is $\approx15\%$, and it is
$\approx22\%$ for our high-luminosity objects.
Scattering alone can provide an $f_{\rm disc}$ value 
of $\sim4\%$, insufficient to power the lines.
Warped accretion discs are probably needed for direct illumination to
work efficiently, especially in high-luminosity objects 
with geometrically thin discs.
See Section 6.1.

\item
We discussed the disc-wind model  
that was proposed
to explain the small fraction of disc-like emitters among 
the general AGN population, and we showed how our current 
results fit into the context of this model.
See Section 6.2.

\item
We discovered a LoBAL quasar, 
J$0850+4451$, among our disc-like emitters.
This object demonstrates that disc-like emission lines and BALs are not exclusive 
phenomena. 
J$0850+4451$ has a relatively small disc inclination angle compared to
traditional BAL quasars.
See Sections 4.2 and 6.2.

\end{enumerate}

Studies of disc-like emitters open a window to probing 
the inner structure 
of AGNs, such 
as the accretion disc, disc wind, and BLR. The high-luminosity and 
relatively high-redshift sample presented here places further constraints on the
disc illumination-mechanism and the corresponding AGN components (accretion
disc and scattering medium). Further spectral 
surveys of X-ray
absorption in disc-like emitters
will help to clarify the covering factor
and optical depth of the
scattering medium. A large and complete
optical survey of disc-like emitters in different luminosity ranges,
utilizing quantitative selection criteria,
will determine whether the fraction of disc-like emitters decreases
as luminosity increases, and thus assess the applicability of the disc-wind
model. It will also be of interest to compare the detailed properties (e.g., SEDs
and X-ray spectra) of 
AGNs with the same luminosity, similar line widths, but different broad
emission-line profiles 
(e.g., single peaked vs. broad shoulders or flat tops), to try to find the underlying physics that determines the shape 
of the line profile.

~\\

We acknowledge financial support from NASA ADP grant NNX10AC99G (BL, WNB),
Chandra X-ray Center grant GO1-12137X (BL, WNB),
and NASA grant NNX08AW82G (WNB, ME). We thank the referee for carefully
reviewing the manuscript and providing helpful comments.

Funding for the SDSS and SDSS-II has been provided by the 
Alfred P. Sloan Foundation,
the Participating Institutions, the National Science Foundation, 
the U.S. Department
of Energy, the National Aeronautics and Space Administration, 
the Japanese Monbukagakusho,
the Max Planck Society, and the Higher Education Funding Council for England. 
The SDSS
Web site is http://www.sdss.org/. The Hobby-Eberly 
Telescope (HET) is a joint project of
the University of Texas at Austin, the Pennsylvania State University, 
Stanford University,
Ludwig-Maximilians-Universit\"at M\"unchen, and Georg-August-Universit\"at 
G\"ottingen. The HET is
named in honor of its principal benefactors, 
William~P.~Hobby and Robert~E.~Eberly.


\begin{thebibliography}{}

\bibitem[\protect\citeauthoryear{{Abazajian} et~al.,}{{Abazajian}
  et~al.}{2009}]{Abazajian2009}
{Abazajian} K.~N.,  et~al., 2009, \apjs, 182, 543

\bibitem[\protect\citeauthoryear{{Abramowicz}, {Igumenshchev}, {Quataert} \&
  {Narayan}}{{Abramowicz} et~al.}{2002}]{Abramowicz2002}
{Abramowicz} M.~A.,  {Igumenshchev} I.~V.,  {Quataert} E.,    {Narayan} R.,
  2002, \apj, 565, 1101

\bibitem[\protect\citeauthoryear{{Alloin}, {Boisson} \& {Pelat}}{{Alloin}
  et~al.}{1988}]{Alloin1988}
{Alloin} D.,  {Boisson} C.,    {Pelat} D.,  1988, \aap, 200, 17

\bibitem[\protect\citeauthoryear{{Arnaud}}{{Arnaud}}{1996}]{Arnaud1996}
{Arnaud} K.~A.,  1996, in {Jacoby} G.~H.,  {Barnes} J.,  eds, Astronomical Data
  Analysis Software and Systems V Vol.~101 of ASP Conf. Ser., {XSPEC: The First
  Ten Years}.
p.~17

\bibitem[\protect\citeauthoryear{{Barvainis}, {Leh{\'a}r}, {Birkinshaw},
  {Falcke} \& {Blundell}}{{Barvainis} et~al.}{2005}]{Barvainis2005}
{Barvainis} R.,  {Leh{\'a}r} J.,  {Birkinshaw} M.,  {Falcke} H.,    {Blundell}
  K.~M.,  2005, \apj, 618, 108

\bibitem[\protect\citeauthoryear{{Becker}, {White} \& {Helfand}}{{Becker}
  et~al.}{1995}]{Becker1995}
{Becker} R.~H.,  {White} R.~L.,    {Helfand} D.~J.,  1995, \apj, 450, 559

\bibitem[\protect\citeauthoryear{{Boksenberg}, {Carswell}, {Allen}, {Fosbury},
  {Penston} \& {Sargent}}{{Boksenberg} et~al.}{1977}]{Boksenberg1977}
{Boksenberg} A.,  {Carswell} R.~F.,  {Allen} D.~A.,  {Fosbury} R.~A.~E.,
  {Penston} M.~V.,    {Sargent} W.~L.~W.,  1977, \mnras, 178, 451

\bibitem[\protect\citeauthoryear{{Boroson} \& {Meyers}}{{Boroson} \&
  {Meyers}}{1992}]{Boroson1992}
{Boroson} T.~A.,  {Meyers} K.~A.,  1992, \apj, 397, 442

\bibitem[\protect\citeauthoryear{{Brandt}, {Mathur} \& {Elvis}}{{Brandt}
  et~al.}{1997}]{Brandt1997}
{Brandt} W.~N.,  {Mathur} S.,    {Elvis} M.,  1997, \mnras, 285, L25

\bibitem[\protect\citeauthoryear{{Broos}, {Townsley}, {Feigelson}, {Getman},
  {Bauer} \& {Garmire}}{{Broos} et~al.}{2010}]{Broos2010}
{Broos} P.~S.,  {Townsley} L.~K.,  {Feigelson} E.~D.,  {Getman} K.~V.,  {Bauer}
  F.~E.,    {Garmire} G.~P.,  2010, \apj, 714, 1582

\bibitem[\protect\citeauthoryear{{Calzetti}, {Armus}, {Bohlin}, {Kinney},
  {Koornneef} \& {Storchi-Bergmann}}{{Calzetti} et~al.}{2000}]{Calzetti2000}
{Calzetti} D.,  {Armus} L.,  {Bohlin} R.~C.,  {Kinney} A.~L.,  {Koornneef} J.,
    {Storchi-Bergmann} T.,  2000, \apj, 533, 682

\bibitem[\protect\citeauthoryear{{Cao} \& {Wang}}{{Cao} \&
  {Wang}}{2006}]{Cao2006}
{Cao} X.,  {Wang} T.-G.,  2006, \apj, 652, 112

\bibitem[\protect\citeauthoryear{{Chajet} \& {Hall}}{{Chajet} \&
  {Hall}}{2012}]{Chajet2012}
{Chajet} L.~S.,  {Hall} P.~B.,  2012, \mnras, submitted

\bibitem[\protect\citeauthoryear{{Chen} \& {Halpern}}{{Chen} \&
  {Halpern}}{1989}]{Chen1989}
{Chen} K.,  {Halpern} J.~P.,  1989, \apj, 344, 115

\bibitem[\protect\citeauthoryear{{Chen}, {Halpern} \& {Filippenko}}{{Chen}
  et~al.}{1989a}]{Chen1989a}
{Chen} K.,  {Halpern} J.~P.,    {Filippenko} A.~V.,  1989a, \apj, 339, 742

\bibitem[\protect\citeauthoryear{{Chen}, {Halpern} \& {Filippenko}}{{Chen}
  et~al.}{1989b}]{Chen1989b}
{Chen} K.,  {Halpern} J.~P.,    {Filippenko} A.~V.,  1989b, \apj, 339, 742

\bibitem[\protect\citeauthoryear{{Collin-Souffrin}}{{Collin-Souffrin}}{1987}]{%
Collin1987}
{Collin-Souffrin} S.,  1987, \aap, 179, 60

\bibitem[\protect\citeauthoryear{{Collin-Souffrin} \&
  {Dumont}}{{Collin-Souffrin} \& {Dumont}}{1989}]{Collin1989}
{Collin-Souffrin} S.,  {Dumont} A.~M.,  1989, \aap, 213, 29

\bibitem[\protect\citeauthoryear{{Condon}, {Cotton}, {Greisen}, {Yin},
  {Perley}, {Taylor} \& {Broderick}}{{Condon} et~al.}{1998}]{Condon1998}
{Condon} J.~J.,  {Cotton} W.~D.,  {Greisen} E.~W.,  {Yin} Q.~F.,  {Perley}
  R.~A.,  {Taylor} G.~B.,    {Broderick} J.~J.,  1998, \aj, 115, 1693

\bibitem[\protect\citeauthoryear{{Corbin}}{{Corbin}}{1995}]{Corbin1995}
{Corbin} M.~R.,  1995, \apj, 447, 496

\bibitem[\protect\citeauthoryear{{Desai} et~al.,}{{Desai}
  et~al.}{2007}]{Desai2007}
{Desai} V.,  et~al., 2007, \apj, 669, 810

\bibitem[\protect\citeauthoryear{{Dickey} \& {Lockman}}{{Dickey} \&
  {Lockman}}{1990}]{Dickey1990}
{Dickey} J.~M.,  {Lockman} F.~J.,  1990, \araa, 28, 215

\bibitem[\protect\citeauthoryear{{Dietrich}, {Hamann}, {Shields}, {Constantin},
  {Vestergaard}, {Chaffee}, {Foltz} \& {Junkkarinen}}{{Dietrich}
  et~al.}{2002}]{Dietrich2002}
{Dietrich} M.,  {Hamann} F.,  {Shields} J.~C.,  {Constantin} A.,  {Vestergaard}
  M.,  {Chaffee} F.,  {Foltz} C.~B.,    {Junkkarinen} V.~T.,  2002, \apj, 581,
  912

\bibitem[\protect\citeauthoryear{{DiPompeo}, {Brotherton} \& {De
  Breuck}}{{DiPompeo} et~al.}{2012}]{DiPompeo2012}
{DiPompeo} M.~A.,  {Brotherton} M.~S.,    {De Breuck} C.,  2012, \apj, 752, 6

\bibitem[\protect\citeauthoryear{{Dopita} \& {Sutherland}}{{Dopita} \&
  {Sutherland}}{1996}]{Dopita1996}
{Dopita} M.~A.,  {Sutherland} R.~S.,  1996, \apjs, 102, 161

\bibitem[\protect\citeauthoryear{{Dumont} \& {Collin-Souffrin}}{{Dumont} \&
  {Collin-Souffrin}}{1990a}]{Dumont1990b}
{Dumont} A.~M.,  {Collin-Souffrin} S.,  1990a, \aap, 229, 313

\bibitem[\protect\citeauthoryear{{Dumont} \& {Collin-Souffrin}}{{Dumont} \&
  {Collin-Souffrin}}{1990b}]{Dumont1990}
{Dumont} A.~M.,  {Collin-Souffrin} S.,  1990b, \aap, 229, 302

\bibitem[\protect\citeauthoryear{{Eracleous} \& {Halpern}}{{Eracleous} \&
  {Halpern}}{1994}]{Eracleous1994}
{Eracleous} M.,  {Halpern} J.~P.,  1994, \apjs, 90, 1

\bibitem[\protect\citeauthoryear{{Eracleous} \& {Halpern}}{{Eracleous} \&
  {Halpern}}{2001}]{Eracleous2001}
{Eracleous} M.,  {Halpern} J.~P.,  2001, \apj, 554, 240

\bibitem[\protect\citeauthoryear{{Eracleous} \& {Halpern}}{{Eracleous} \&
  {Halpern}}{2003}]{Eracleous2003}
{Eracleous} M.,  {Halpern} J.~P.,  2003, \apj, 599, 886

\bibitem[\protect\citeauthoryear{{Eracleous}, {Halpern} \&
  {Charlton}}{{Eracleous} et~al.}{2003}]{Eracleous2003a}
{Eracleous} M.,  {Halpern} J.~P.,    {Charlton} J.~C.,  2003, \apj, 582, 633

\bibitem[\protect\citeauthoryear{{Eracleous}, {Halpern}, {Storchi-Bergmann},
  {Filippenko}, {Wilson} \& {Livio}}{{Eracleous} et~al.}{2004}]{Eracleous2004}
{Eracleous} M.,  {Halpern} J.~P.,  {Storchi-Bergmann} T.,  {Filippenko} A.~V.,
  {Wilson} A.~S.,    {Livio} M.,  2004, in {Storchi-Bergmann} T.,  {Ho} L.~C.,
   {Schmitt} H.~R.,  eds, The Interplay Among Black Holes, Stars and ISM in
  Galactic Nuclei Vol.~222 of IAU Symposium.
p.~29

\bibitem[\protect\citeauthoryear{{Eracleous}, {Lewis} \& {Flohic}}{{Eracleous}
  et~al.}{2009}]{Eracleous2009}
{Eracleous} M.,  {Lewis} K.~T.,    {Flohic} H.~M.~L.~G.,  2009, \nar, 53, 133

\bibitem[\protect\citeauthoryear{{Eracleous}, {Livio}, {Halpern} \&
  {Storchi-Bergmann}}{{Eracleous} et~al.}{1995}]{Eracleous1995}
{Eracleous} M.,  {Livio} M.,  {Halpern} J.~P.,    {Storchi-Bergmann} T.,  1995,
  \apj, 438, 610

\bibitem[\protect\citeauthoryear{{Falcke}, {Sherwood} \& {Patnaik}}{{Falcke}
  et~al.}{1996}]{Falcke1996}
{Falcke} H.,  {Sherwood} W.,    {Patnaik} A.~R.,  1996, \apj, 471, 106

\bibitem[\protect\citeauthoryear{{Flohic}, {Eracleous} \&
  {Bogdanovi{\'c}}}{{Flohic} et~al.}{2012}]{Flohic2012}
{Flohic} H.~M.~L.~G.,  {Eracleous} M.,    {Bogdanovi{\'c}} T.,  2012, \apj,
  753, 133

\bibitem[\protect\citeauthoryear{{Freeman}, {Kashyap}, {Rosner} \&
  {Lamb}}{{Freeman} et~al.}{2002}]{Freeman2002}
{Freeman} P.~E.,  {Kashyap} V.,  {Rosner} R.,    {Lamb} D.~Q.,  2002, \apjs,
  138, 185

\bibitem[\protect\citeauthoryear{{Gallagher}, {Brandt}, {Chartas} \&
  {Garmire}}{{Gallagher} et~al.}{2002}]{Gallagher2002}
{Gallagher} S.~C.,  {Brandt} W.~N.,  {Chartas} G.,    {Garmire} G.~P.,  2002,
  \apj, 567, 37

\bibitem[\protect\citeauthoryear{{Gallagher}, {Brandt}, {Chartas}, {Priddey},
  {Garmire} \& {Sambruna}}{{Gallagher} et~al.}{2006}]{Gallagher2006}
{Gallagher} S.~C.,  {Brandt} W.~N.,  {Chartas} G.,  {Priddey} R.,  {Garmire}
  G.~P.,    {Sambruna} R.~M.,  2006, \apj, 644, 709

\bibitem[\protect\citeauthoryear{{Gallagher}, {Hines}, {Blaylock}, {Priddey},
  {Brandt} \& {Egami}}{{Gallagher} et~al.}{2007}]{Gallagher2007}
{Gallagher} S.~C.,  {Hines} D.~C.,  {Blaylock} M.,  {Priddey} R.~S.,  {Brandt}
  W.~N.,    {Egami} E.~E.,  2007, \apj, 665, 157

\bibitem[\protect\citeauthoryear{{Ganguly}, {Brotherton}, {Cales}, {Scoggins},
  {Shang} \& {Vestergaard}}{{Ganguly} et~al.}{2007}]{Ganguly2007}
{Ganguly} R.,  {Brotherton} M.~S.,  {Cales} S.,  {Scoggins} B.,  {Shang} Z.,
  {Vestergaard} M.,  2007, \apj, 665, 990

\bibitem[\protect\citeauthoryear{{Garmire}, {Bautz}, {Ford}, {Nousek} \&
  {Ricker} Jr.}{{Garmire} et~al.}{2003}]{Garmire2003}
{Garmire} G.~P.,  {Bautz} M.~W.,  {Ford} P.~G.,  {Nousek} J.~A.,    {Ricker}
  Jr. G.~R.,  2003, Proc. SPIE, 4851, 28

\bibitem[\protect\citeauthoryear{{Gaskell}}{{Gaskell}}{2009}]{Gaskell2009}
{Gaskell} C.~M.,  2009, \nar, 53, 140

\bibitem[\protect\citeauthoryear{{Gehrels}}{{Gehrels}}{1986}]{Gehrels1986}
{Gehrels} N.,  1986, \apj, 303, 336

\bibitem[\protect\citeauthoryear{{Gezari}, {Halpern} \& {Eracleous}}{{Gezari}
  et~al.}{2007}]{Gezari2007}
{Gezari} S.,  {Halpern} J.~P.,    {Eracleous} M.,  2007, \apjs, 169, 167

\bibitem[\protect\citeauthoryear{{Gibson} \& {Brandt}}{{Gibson} \&
  {Brandt}}{2012}]{Gibson2012}
{Gibson} R.~R.,  {Brandt} W.~N.,  2012, \apj, 746, 54

\bibitem[\protect\citeauthoryear{{Gibson} et~al.,}{{Gibson}
  et~al.}{2009}]{Gibson2009}
{Gibson} R.~R.,  et~al., 2009, \apj, 692, 758

\bibitem[\protect\citeauthoryear{{Green}, {Aldcroft}, {Mathur}, {Wilkes} \&
  {Elvis}}{{Green} et~al.}{2001}]{Green2001}
{Green} P.~J.,  {Aldcroft} T.~L.,  {Mathur} S.,  {Wilkes} B.~J.,    {Elvis} M.,
   2001, \apj, 558, 109

\bibitem[\protect\citeauthoryear{{Green} et~al.,}{{Green}
  et~al.}{1995}]{Green1995}
{Green} P.~J.,  et~al., 1995, \apj, 450, 51

\bibitem[\protect\citeauthoryear{{Greenstein} \& {Kraft}}{{Greenstein} \&
  {Kraft}}{1959}]{Greenstein1959}
{Greenstein} J.~L.,  {Kraft} R.~P.,  1959, \apj, 130, 99

\bibitem[\protect\citeauthoryear{{Halpern} \& {Eracleous}}{{Halpern} \&
  {Eracleous}}{1994}]{Halpern1994}
{Halpern} J.~P.,  {Eracleous} M.,  1994, \apjl, 433, L17

\bibitem[\protect\citeauthoryear{{Halpern}, {Eracleous}, {Filippenko} \&
  {Chen}}{{Halpern} et~al.}{1996}]{Halpern1996}
{Halpern} J.~P.,  {Eracleous} M.,  {Filippenko} A.~V.,    {Chen} K.,  1996,
  \apj, 464, 704

\bibitem[\protect\citeauthoryear{{Hao}, {Elvis}, {Civano} \& {Lawrence}}{{Hao}
  et~al.}{2011}]{Hao2011}
{Hao} H.,  {Elvis} M.,  {Civano} F.,    {Lawrence} A.,  2011, \apj, 733, 108

\bibitem[\protect\citeauthoryear{{Hao} et~al.,}{{Hao}  et~al.}{2010}]{Hao2010}
{Hao} H.,  et~al., 2010, \apjl, 724, L59

\bibitem[\protect\citeauthoryear{{Hewett} \& {Foltz}}{{Hewett} \&
  {Foltz}}{2003}]{Hewett2003}
{Hewett} P.~C.,  {Foltz} C.~B.,  2003, \aj, 125, 1784

\bibitem[\protect\citeauthoryear{{Hewett} \& {Wild}}{{Hewett} \&
  {Wild}}{2010}]{Hewett2010}
{Hewett} P.~C.,  {Wild} V.,  2010, \mnras, 405, 2302

\bibitem[\protect\citeauthoryear{{Hill}, {Nicklas}, {MacQueen}, {Tejada},
  {Cobos Duenas} \& {Mitsch}}{{Hill} et~al.}{1998}]{Hill1998}
{Hill} G.~J.,  {Nicklas} H.~E.,  {MacQueen} P.~J.,  {Tejada} C.,  {Cobos
  Duenas} F.~J.,    {Mitsch} W.,  1998, Proc. SPIE, 3355, 375

\bibitem[\protect\citeauthoryear{{Jackson}, {Perez} \& {Penston}}{{Jackson}
  et~al.}{1991}]{Jackson1991}
{Jackson} N.,  {Perez} E.,    {Penston} M.~V.,  1991, \mnras, 249, 577

\bibitem[\protect\citeauthoryear{{Jiang}, {Wang} \& {Wang}}{{Jiang}
  et~al.}{2006}]{Jiang2006}
{Jiang} P.,  {Wang} J.~X.,    {Wang} T.~G.,  2006, \apj, 644, 725

\bibitem[\protect\citeauthoryear{{Just}, {Brandt}, {Shemmer}, {Steffen},
  {Schneider}, {Chartas} \& {Garmire}}{{Just} et~al.}{2007}]{Just2007}
{Just} D.~W.,  {Brandt} W.~N.,  {Shemmer} O.,  {Steffen} A.~T.,  {Schneider}
  D.~P.,  {Chartas} G.,    {Garmire} G.~P.,  2007, \apj, 665, 1004

\bibitem[\protect\citeauthoryear{{Kellermann}, {Sramek}, {Schmidt}, {Shaffer}
  \& {Green}}{{Kellermann} et~al.}{1989}]{Kellermann1989}
{Kellermann} K.~I.,  {Sramek} R.,  {Schmidt} M.,  {Shaffer} D.~B.,    {Green}
  R.,  1989, \aj, 98, 1195

\bibitem[\protect\citeauthoryear{{Komatsu} et~al.,}{{Komatsu}
  et~al.}{2011}]{Komatsu2011}
{Komatsu} E.,  et~al., 2011, \apjs, 192, 18

\bibitem[\protect\citeauthoryear{{Korista}, {Baldwin} \& {Ferland}}{{Korista}
  et~al.}{1998}]{Korista1998}
{Korista} K.,  {Baldwin} J.,    {Ferland} G.,  1998, \apj, 507, 24

\bibitem[\protect\citeauthoryear{{Kraft}, {Burrows} \& {Nousek}}{{Kraft}
  et~al.}{1991}]{Kraft1991}
{Kraft} R.~P.,  {Burrows} D.~N.,    {Nousek} J.~A.,  1991, \apj, 374, 344

\bibitem[\protect\citeauthoryear{{Lawrence} \& {Elvis}}{{Lawrence} \&
  {Elvis}}{2010}]{Lawrence2010}
{Lawrence} A.,  {Elvis} M.,  2010, \apj, 714, 561

\bibitem[\protect\citeauthoryear{{Leighly}, {Dietrich} \& {Barber}}{{Leighly}
  et~al.}{2011}]{Leighly2011}
{Leighly} K.~M.,  {Dietrich} M.,    {Barber} S.,  2011, \apj, 728, 94

\bibitem[\protect\citeauthoryear{{Lewis} \& {Eracleous}}{{Lewis} \&
  {Eracleous}}{2006}]{Lewis2006}
{Lewis} K.~T.,  {Eracleous} M.,  2006, \apj, 642, 711

\bibitem[\protect\citeauthoryear{{Lewis}, {Eracleous} \&
  {Storchi-Bergmann}}{{Lewis} et~al.}{2010}]{Lewis2010}
{Lewis} K.~T.,  {Eracleous} M.,    {Storchi-Bergmann} T.,  2010, \apjs, 187,
  416

\bibitem[\protect\citeauthoryear{{Luo} et~al.,}{{Luo}  et~al.}{2009}]{Luo2009}
{Luo} B.,  et~al., 2009, \apj, 695, 1227

\bibitem[\protect\citeauthoryear{{MacAlpine}}{{MacAlpine}}{2003}]{MacAlpine200%
3}
{MacAlpine} G.~M.,  2003, in {Reyes-Ruiz} M.,  {V{\'a}zquez-Semadeni} E.,  eds,
  Revista Mexicana de Astronomia y Astrofisica Conference Series Vol.~18 of
  Revista Mexicana de Astronomia y Astrofisica Conference Series.
p.~63

\bibitem[\protect\citeauthoryear{{Maiolino}, {Salvati}, {Marconi} \&
  {Antonucci}}{{Maiolino} et~al.}{2001}]{Maiolino2001}
{Maiolino} R.,  {Salvati} M.,  {Marconi} A.,    {Antonucci} R.~R.~J.,  2001,
  \aap, 375, 25

\bibitem[\protect\citeauthoryear{{Maloney}, {Begelman} \& {Pringle}}{{Maloney}
  et~al.}{1996}]{Maloney1996}
{Maloney} P.~R.,  {Begelman} M.~C.,    {Pringle} J.~E.,  1996, \apj, 472, 582

\bibitem[\protect\citeauthoryear{{Maoz}}{{Maoz}}{2007}]{Maoz2007}
{Maoz} D.,  2007, \mnras, 377, 1696

\bibitem[\protect\citeauthoryear{{Marsh}}{{Marsh}}{1988}]{Marsh1988}
{Marsh} T.~R.,  1988, \mnras, 231, 1117

\bibitem[\protect\citeauthoryear{{Martin} et~al.,}{{Martin}
  et~al.}{2005}]{Martin2005}
{Martin} D.~C.,  et~al., 2005, \apjl, 619, L1

\bibitem[\protect\citeauthoryear{Miller}{Miller}{}]{Miller2011}
Miller

\bibitem[\protect\citeauthoryear{{Murray} \& {Chiang}}{{Murray} \&
  {Chiang}}{1997}]{Murray1997}
{Murray} N.,  {Chiang} J.,  1997, \apj, 474, 91

\bibitem[\protect\citeauthoryear{{Murray}, {Chiang}, {Grossman} \&
  {Voit}}{{Murray} et~al.}{1995}]{Murray1995}
{Murray} N.,  {Chiang} J.,  {Grossman} S.~A.,    {Voit} G.~M.,  1995, \apj,
  451, 498

\bibitem[\protect\citeauthoryear{{Nandra}, {Georgantopoulos}, {Ptak} \&
  {Turner}}{{Nandra} et~al.}{2003}]{Nandra2003}
{Nandra} K.,  {Georgantopoulos} I.,  {Ptak} A.,    {Turner} T.~J.,  2003, \apj,
  582, 615

\bibitem[\protect\citeauthoryear{{Narayan} \& {Yi}}{{Narayan} \&
  {Yi}}{1994}]{Narayan1994}
{Narayan} R.,  {Yi} I.,  1994, \apjl, 428, L13

\bibitem[\protect\citeauthoryear{{Ogle}, {Cohen}, {Miller}, {Tran}, {Goodrich}
  \& {Martel}}{{Ogle} et~al.}{1999}]{Ogle1999}
{Ogle} P.~M.,  {Cohen} M.~H.,  {Miller} J.~S.,  {Tran} H.~D.,  {Goodrich}
  R.~W.,    {Martel} A.~R.,  1999, \apjs, 125, 1

\bibitem[\protect\citeauthoryear{{Oke}}{{Oke}}{1987}]{Oke1987}
{Oke} J.~B.,  1987, in {Zensus} J.~A.,  {Pearson} T.~J.,  eds, Superluminal
  Radio Sources {Emission-line profile changes in 3C 390.3}.
p.~267

\bibitem[\protect\citeauthoryear{{Peterson} et~al.,}{{Peterson}
  et~al.}{1999}]{Peterson1999}
{Peterson} B.~M.,  et~al., 1999, \apj, 510, 659

\bibitem[\protect\citeauthoryear{{Primini} et~al.,}{{Primini}
  et~al.}{2011}]{Primini2011}
{Primini} F.~A.,  et~al., 2011, \apjs, 194, 37

\bibitem[\protect\citeauthoryear{{Pringle}}{{Pringle}}{1996}]{Pringle1996}
{Pringle} J.~E.,  1996, \mnras, 281, 357

\bibitem[\protect\citeauthoryear{{Rafferty}, {Brandt}, {Alexander}, {Xue},
  {Bauer}, {Lehmer}, {Luo} \& {Papovich}}{{Rafferty}
  et~al.}{2011}]{Rafferty2011}
{Rafferty} D.~A.,  {Brandt} W.~N.,  {Alexander} D.~M.,  {Xue} Y.~Q.,  {Bauer}
  F.~E.,  {Lehmer} B.~D.,  {Luo} B.,    {Papovich} C.,  2011, \apj, 742, 3

\bibitem[\protect\citeauthoryear{{Ramsey} et~al.,}{{Ramsey}
  et~al.}{1998}]{Ramsey1998}
{Ramsey} L.~W.,  et~al., 1998, Proc. SPIE, 3352, 34

\bibitem[\protect\citeauthoryear{{Rees}, {Begelman}, {Blandford} \&
  {Phinney}}{{Rees} et~al.}{1982}]{Rees1982}
{Rees} M.~J.,  {Begelman} M.~C.,  {Blandford} R.~D.,    {Phinney} E.~S.,  1982,
  \nat, 295, 17

\bibitem[\protect\citeauthoryear{{Reeves} \& {Turner}}{{Reeves} \&
  {Turner}}{2000}]{Reeves2000}
{Reeves} J.~N.,  {Turner} M.~J.~L.,  2000, \mnras, 316, 234

\bibitem[\protect\citeauthoryear{{Reeves}, {Turner}, {Ohashi} \&
  {Kii}}{{Reeves} et~al.}{1997}]{Reeves1997}
{Reeves} J.~N.,  {Turner} M.~J.~L.,  {Ohashi} T.,    {Kii} T.,  1997, \mnras,
  292, 468

\bibitem[\protect\citeauthoryear{{Reyes} et~al.,}{{Reyes}
  et~al.}{2008}]{Reyes2008}
{Reyes} R.,  et~al., 2008, \aj, 136, 2373

\bibitem[\protect\citeauthoryear{{Richards} et~al.,}{{Richards}
  et~al.}{2006}]{Richards2006}
{Richards} G.~T.,  et~al., 2006, \apjs, 166, 470

\bibitem[\protect\citeauthoryear{{Rokaki}, {Boisson} \&
  {Collin-Souffrin}}{{Rokaki} et~al.}{1992}]{Rokaki1992}
{Rokaki} E.,  {Boisson} C.,    {Collin-Souffrin} S.,  1992, \aap, 253, 57

\bibitem[\protect\citeauthoryear{{Schneider} et~al.,}{{Schneider}
  et~al.}{2010}]{Schneider2010}
{Schneider} D.~P.,  et~al., 2010, \aj, 139, 2360

\bibitem[\protect\citeauthoryear{{Scott}, {Stewart}, {Mateos}, {Alexander},
  {Hutton} \& {Ward}}{{Scott} et~al.}{2011}]{Scott2011}
{Scott} A.~E.,  {Stewart} G.~C.,  {Mateos} S.,  {Alexander} D.~M.,  {Hutton}
  S.,    {Ward} M.~J.,  2011, \mnras, 417, 992

\bibitem[\protect\citeauthoryear{{Sergeev}, {Doroshenko}, {Dzyuba}, {Peterson},
  {Pogge} \& {Pronik}}{{Sergeev} et~al.}{2007}]{Sergeev2007}
{Sergeev} S.~G.,  {Doroshenko} V.~T.,  {Dzyuba} S.~A.,  {Peterson} B.~M.,
  {Pogge} R.~W.,    {Pronik} V.~I.,  2007, \apj, 668, 708

\bibitem[\protect\citeauthoryear{{Shakura} \& {Sunyaev}}{{Shakura} \&
  {Sunyaev}}{1973}]{Shakura1973}
{Shakura} N.~I.,  {Sunyaev} R.~A.,  1973, \aap, 24, 337

\bibitem[\protect\citeauthoryear{{Shen} et~al.,}{{Shen}
  et~al.}{2011}]{Shen2011}
{Shen} Y.,  et~al., 2011, \apjs, 194, 45

\bibitem[\protect\citeauthoryear{{Shi} et~al.,}{{Shi}  et~al.}{2007}]{Shi2007}
{Shi} Y.,  et~al., 2007, \apj, 669, 841

\bibitem[\protect\citeauthoryear{{Skrutskie} et~al.,}{{Skrutskie}
  et~al.}{2006}]{Skrutskie2006}
{Skrutskie} M.~F.,  et~al., 2006, \aj, 131, 1163

\bibitem[\protect\citeauthoryear{{Stauffer}, {Schild} \& {Keel}}{{Stauffer}
  et~al.}{1983}]{Stauffer1983}
{Stauffer} J.,  {Schild} R.,    {Keel} W.,  1983, \apj, 270, 465

\bibitem[\protect\citeauthoryear{{Steffen}, {Strateva}, {Brandt}, {Alexander},
  {Koekemoer}, {Lehmer}, {Schneider} \& {Vignali}}{{Steffen}
  et~al.}{2006}]{Steffen2006}
{Steffen} A.~T.,  {Strateva} I.,  {Brandt} W.~N.,  {Alexander} D.~M.,
  {Koekemoer} A.~M.,  {Lehmer} B.~D.,  {Schneider} D.~P.,    {Vignali} C.,
  2006, \aj, 131, 2826

\bibitem[\protect\citeauthoryear{{Stern} \& {Laor}}{{Stern} \&
  {Laor}}{2012}]{Stern2012}
{Stern} J.,  {Laor} A.,  2012, \mnras, 423, 600

\bibitem[\protect\citeauthoryear{{Storchi-Bergmann} et~al.,}{{Storchi-Bergmann}
   et~al.}{2003}]{Storchi2003}
{Storchi-Bergmann} T.,  et~al., 2003, \apj, 598, 956

\bibitem[\protect\citeauthoryear{{Strateva}, {Brandt}, {Eracleous} \&
  {Garmire}}{{Strateva} et~al.}{2008}]{Strateva2008}
{Strateva} I.~V.,  {Brandt} W.~N.,  {Eracleous} M.,    {Garmire} G.,  2008,
  \apj, 687, 869

\bibitem[\protect\citeauthoryear{{Strateva}, {Brandt}, {Eracleous}, {Schneider}
  \& {Chartas}}{{Strateva} et~al.}{2006}]{Strateva2006}
{Strateva} I.~V.,  {Brandt} W.~N.,  {Eracleous} M.,  {Schneider} D.~P.,
  {Chartas} G.,  2006, \apj, 651, 749

\bibitem[\protect\citeauthoryear{{Strateva} et~al.,}{{Strateva}
  et~al.}{2003}]{Strateva2003}
{Strateva} I.~V.,  et~al., 2003, \aj, 126, 1720

\bibitem[\protect\citeauthoryear{{Sulentic}, {Marziani}, {Zwitter} \&
  {Calvani}}{{Sulentic} et~al.}{1995}]{Sulentic1995}
{Sulentic} J.~W.,  {Marziani} P.,  {Zwitter} T.,    {Calvani} M.,  1995, \apjl,
  438, 1

\bibitem[\protect\citeauthoryear{{Trump} et~al.,}{{Trump}
  et~al.}{2006}]{Trump2006}
{Trump} J.~R.,  et~al., 2006, \apjs, 165, 1

\bibitem[\protect\citeauthoryear{{Vanden Berk} et~al.,}{{Vanden Berk}
  et~al.}{2001}]{Vandenberk2001}
{Vanden Berk} D.~E.,  et~al., 2001, \aj, 122, 549

\bibitem[\protect\citeauthoryear{{V{\'e}ron-Cetty}, {Joly} \&
  {V{\'e}ron}}{{V{\'e}ron-Cetty} et~al.}{2004}]{Veron2004}
{V{\'e}ron-Cetty} M.,  {Joly} M.,    {V{\'e}ron} P.,  2004, \aap, 417, 515

\bibitem[\protect\citeauthoryear{{Vestergaard} \& {Wilkes}}{{Vestergaard} \&
  {Wilkes}}{2001}]{Vestergaard2001}
{Vestergaard} M.,  {Wilkes} B.~J.,  2001, \apjs, 134, 1

\bibitem[\protect\citeauthoryear{{Weymann}, {Morris}, {Foltz} \&
  {Hewett}}{{Weymann} et~al.}{1991}]{Weymann1991}
{Weymann} R.~J.,  {Morris} S.~L.,  {Foltz} C.~B.,    {Hewett} P.~C.,  1991,
  \apj, 373, 23

\bibitem[\protect\citeauthoryear{{White}, {Becker}, {Helfand} \&
  {Gregg}}{{White} et~al.}{1997}]{White1997}
{White} R.~L.,  {Becker} R.~H.,  {Helfand} D.~J.,    {Gregg} M.~D.,  1997,
  \apj, 475, 479

\bibitem[\protect\citeauthoryear{{Williams}}{{Williams}}{1980}]{Williams1980}
{Williams} R.~E.,  1980, \apj, 235, 939

\bibitem[\protect\citeauthoryear{{Worrall}, {Tananbaum}, {Giommi} \&
  {Zamorani}}{{Worrall} et~al.}{1987}]{Worrall1987}
{Worrall} D.~M.,  {Tananbaum} H.,  {Giommi} P.,    {Zamorani} G.,  1987, \apj,
  313, 596

\bibitem[\protect\citeauthoryear{{Wright} et~al.,}{{Wright}
  et~al.}{2010}]{Wright2010}
{Wright} E.~L.,  et~al., 2010, \aj, 140, 1868

\bibitem[\protect\citeauthoryear{{Wu}, {Vanden Berk}, {Brandt}, {Schneider},
  {Gibson} \& {Wu}}{{Wu} et~al.}{2009}]{Wu2009}
{Wu} J.,  {Vanden Berk} D.~E.,  {Brandt} W.~N.,  {Schneider} D.~P.,  {Gibson}
  R.~R.,    {Wu} J.,  2009, \apj, 702, 767

\bibitem[\protect\citeauthoryear{{Wu}, {Wang} \& {Dong}}{{Wu}
  et~al.}{2008}]{Wu2008}
{Wu} S.-M.,  {Wang} T.-G.,    {Dong} X.-B.,  2008, \mnras, 389, 213

\bibitem[\protect\citeauthoryear{{York} et~al.,}{{York}
  et~al.}{2000}]{York2000}
{York} D.~G.,  et~al., 2000, \aj, 120, 1579

\bibitem[\protect\citeauthoryear{{Young} \& {Schneider}}{{Young} \&
  {Schneider}}{1980}]{Young1980}
{Young} P.,  {Schneider} D.~P.,  1980, \apj, 238, 955

\bibitem[\protect\citeauthoryear{{Young}, {Schneider} \& {Shectman}}{{Young}
  et~al.}{1981}]{Young1981}
{Young} P.,  {Schneider} D.~P.,    {Shectman} S.~A.,  1981, \apj, 245, 1035

\end{thebibliography}

%

\begin{table*}
\centering
\begin{minipage}{140mm}
\caption{X-ray Observation Log}
\begin{tabular}{cccccccc}
\hline
{Object Name}                   &
{$z$}                   &
{$\Delta_{\rm OX}$}                   &
{Observation}                   &
{Observation}                   &
{Exp}                   &
{Exp\_clean}                   &
{Off-axis Angle}  \\
{(SDSS J)}                   &
{}                   &
{(arcsec)}                   &
{Date}                   &
{ID}                   &
{(ks)}                   &
{(ks)}                   &
{(arcmin)}  \\
{(1)}         &
{(2)}         &
{(3)}         &
{(4)}         &
{(5)}         &
{(6)}         &
{(7)}         &
{(8)}         \\

\hline
$     040148.98-054056.5$&0.571&0.1&         2011 Aug 27&  12803&   5.0&   4.3&$  0.2$\\
$     085053.12+445122.4$&0.542&0.2&         2011 Dec 29&  12807&   7.9&   7.4&$  0.2$\\
$     095948.59+344915.7$&0.706&0.1&         2012 Jan 11&  12806&  11.0&  10.4&$  0.2$\\
$     150017.58+121036.5$&0.783&0.3&         2011 May 29&  12802&   5.0&   4.7&$  0.3$\\
$     153159.10+242047.1$&0.632&0.1&         2002 Sep 25&   3336&   5.2&   5.0&$  1.1$\\
$     160918.93+082416.4$&0.732&0.3&         2011 Dec 29&  12804&  10.0&   9.4&$  0.3$\\
$     212501.20-081328.6$&0.625&0.2&         2008 Jul 10&   9183&  40.2&  36.2&$  0.3$\\
$     214843.56+001054.5$&0.673&0.2&         2011 May 20&  12805&   9.0&   8.5&$  0.3$\\
\hline
\end{tabular}

Note. --- Col. (1): SDSS name. The sources are listed in order of increasing right ascension. J$1531+2420$ and J$2125-0813$ have archival \chandra\ data, and the others are Cycle 12 targets.
Col. (2): redshift from \citet{Hewett2010}.
Col. (3): angular distance between optical and X-ray positions.
Cols. (4) and (5): \chandra\ observation date and ID.
Col. (6): \chandra\ exposure time.
Col. (7): Background-flare cleaned effective exposure time in the 0.5--8.0 keV band at the source position.
Col. (8): off-axis angle of the source, defined as the distance between the source and the observation aim point.
\label{tbl-log}
\end{minipage}
\end{table*}

%
%

\begin{table*}
\centering
\begin{minipage}{170mm}
\caption{H$\beta$ Line Disc-Fit Parameters}
\begin{tabular}{ccccccccccccc}
\hline
{Object Name} &
{FWHM}                   &
{$R_{\rm in}$}                   &
{$R_{\rm out}$}                   &
{$i$}                   &
{$q_1$}                   &
{$q_2$}                   &
{$R_{\rm b}$}     &
{$\sigma$}     &
{$\log L_{\rm H\beta}$} &
{$\log L_{\rm bol}$} &
{$\log W_{\rm d}$} &
{$L_{\rm H\beta}/W_{\rm d}$} \\
{}   &
{(km s$^{-1}$)}                   &
{($R_{\rm G}$)}                   &
{($R_{\rm G}$)}                   &
{(deg)}                   &
{}                   &
{}                   &
{($R_{\rm G}$)}     &
{(km s$^{-1}$)}     &
{(\lum)}     &
{(\lum)}     &
{(\lum)}     &
{}     \\
{(1)}         &
{(2)}         &
{(3)}         &
{(4)}         &
{(5)}         &
{(6)}         &
{(7)}         &
{(8)}         &
{(9)}         &
{(10)}         &
{(11)}         &
{(12)}         &
{(13)}         \\
\hline
{            J$0401-0540$}& 7200&  120&  3700&20& 1.8&...&  ...&   900&  43.9&  46.7&  45.4& 0.03\\
{            J$0850+4451$}& 9300&  450&  7000&34& 2.1&...&  ...&  1000&  43.4&  46.1&  44.3& 0.13\\
{            J$0959+3449$}& 9400&  110&  2300&22& 1.8&...&  ...&  1300&  43.6&  46.2&  45.0& 0.05\\
{            J$1500+1210$}& 6400&  100&  5000&19& 2.0&1.7&  900&   950&  44.5&  47.2&  46.0& 0.03\\
{            J$1531+2420$}&11100&  110&  3500&22& 2.4&...&  ...&  1000&  43.7&  46.7&  45.4& 0.02\\
{            J$1609+0824$}& 7300&  130&  2800&18& 2.3&  0& 1500&   800&  43.7&  46.4&  45.1& 0.04\\
{            J$2125-0813$}& 8600&  230&  5000&29& 1.8&...&  ...&   800&  43.6&  46.5&  45.0& 0.05\\
{            J$2148+0010$}&11200&  250&  5000&33& 2.3&...&  ...&  1100&  43.8&  46.5&  45.0& 0.07\\
\hline
\end{tabular}

Note. --- Col. (1): abbreviated SDSS name.
Col. (2): FWHM of the broad H$\beta$ line.
Col. (3): inner radius of the disc-emission region.
Col. (4): outer radius of the disc-emission region.
Col. (5): disc inclination angle.
Col. (6): power-law index of the disc emissivity. A simple power-law or broken power-law model was used.
Col. (7): power-law index of the disc emissivity outside the break radius, when adopting a broken power-law model.
Col. (8): break radius, when adopting a broken power-law model.
Col. (9): broadening parameter.
Col. (10): integrated luminosity of the disc-like H$\beta$ line.
Col. (11): bolometric luminosity estimated based on the \citet{Richards2006} SED templates (see Fig.~\ref{fig-sed}).
Col. (12): viscous power available locally in the emission region of the accretion disc.
Col. (12): ratio of the H$\beta$ line luminosity to the local viscous power of the disc.
\label{tbl-fit}
\end{minipage}
\end{table*}

%
%

\begin{table*}
\centering
\begin{minipage}{155mm}

\caption{X-ray Counts and Spectral Properties}
\begin{tabular}{cccccccc}
\hline
{Object Name} &
{Full Band} &
{Soft Band} &
{Hard Band} &
{$N_{\rm H,Gal}$}    &
{$N_{\rm H,int}$}  &
{$\Gamma$}                   &
{$\chi^2/$dof}  \\
{}                   &
{(0.5--8.0 keV)}                   &
{(0.5--2.0 keV)}                   &
{(2.0--8.0 keV)}                   &
{($\times10^{20}~{\rm cm}^{-2}$)}                   &
{($\times10^{22}~{\rm cm}^{-2}$)}                   &
{ }                   &
{ }                   \\
{(1)}         &
{(2)}         &
{(3)}         &
{(4)}         &
{(5)}         &
{(6)}         &
{(7)}         &
{(8)}         \\

\hline
            J$0401-0540$&$  401.5_{-21.3}^{+22.5}$&$  309.7_{-18.3}^{+19.4}$&$   84.7_{-10.2}^{+11.5}$& 6.8&$                  <0.3$&$     2.0_{-0.1}^{+0.2}$&30.0/33\\
            J$0850+4451$&$      2.8_{-1.7}^{+3.1}$&$                   <4.7$&$      3.1_{-1.8}^{+3.2}$& 2.6&$                   ...$&$                   ...$&    ...\\
            J$0959+3449$&$  550.0_{-24.9}^{+26.1}$&$  415.0_{-21.2}^{+22.3}$&$  126.5_{-12.4}^{+13.7}$& 1.3&$                 <0.08$&$     1.9_{-0.1}^{+0.1}$&37.4/33\\
            J$1500+1210$&$  273.6_{-17.6}^{+18.8}$&$  156.9_{-13.0}^{+14.2}$&$  119.1_{-12.1}^{+13.4}$& 2.4&$     0.8_{-0.6}^{+0.8}$&$     1.4_{-0.4}^{+0.4}$& 8.3/22\\
            J$1531+2420$&$  769.5_{-29.6}^{+30.7}$&$  565.6_{-24.8}^{+25.9}$&$  194.4_{-15.5}^{+16.8}$& 4.1&$                 <0.02$&$     1.9_{-0.1}^{+0.1}$&50.7/46\\
            J$1609+0824$&$  859.9_{-31.2}^{+32.3}$&$  651.5_{-26.6}^{+27.7}$&$  195.1_{-15.5}^{+16.7}$& 4.0&$                 <0.06$&$     1.9_{-0.1}^{+0.1}$&48.9/48\\
            J$2125-0813$&$ 3215.0_{-60.4}^{+61.5}$&$ 2036.9_{-47.0}^{+48.1}$&$ 1180.9_{-38.1}^{+39.4}$& 5.6&$                 <0.02$&$  1.39_{-0.05}^{+0.05}$&148/157\\
            J$2148+0010$&$   88.1_{-10.0}^{+11.2}$&$     31.4_{-5.8}^{+7.0}$&$     60.1_{-8.6}^{+9.9}$& 5.9&$     2.5_{-1.0}^{+1.2}$&$                   1.8$&  78/80\\
\hline
\end{tabular}

Note. --- Col. (1): abbreviated SDSS name.
Cols. (2)--(4): X-ray counts in the full, soft, and hard bands.
Col. (5): Galactic neutral hydrogen column density \citep{Dickey1990}.
Cols. (6)--(8): best-fitting spectral model from XSPEC fitting. We adopted an absorbed power-law model ({\sc wabs*zpow*zwabs}). In cases where intrinsic absorption is not required, we adopted a power-law model ({\sc wabs*zpow}) and the 90\% confidence upper limits on the intrinsic hydrogen column densities are given. For J2148$+$0010, the $C$ statistic (cstat) was used in XSPEC due to this source's limited number of source counts, and we fixed the photon index at $\Gamma=1.8$. The $C$-statistic value divided by the number of degrees of freedom is reported in col. (8) for this source. We did not attempt to fit the spectrum of J0850$+$4451 owing to the very small number of counts.
\label{tbl-xspec}
\end{minipage}
\end{table*}

%
%

\begin{table*}
\centering
\begin{minipage}{158mm}

\caption{X-ray, Optical, and Radio Properties}
\begin{tabular}{cccccccccc}
\hline

{Object Name} &
{$M_i$}                   &
{Count}                   &
{$f_{\rm 2 keV}$}                   &
{$\log L_{\rm X}$}                   &
{$f_{\rm 2500 \AA}$}                   &
{$\log L_{\rm 2500 \AA}$}                   &
{$\alpha_{\rm OX}$}     &
{$\Delta\alpha_{\rm OX}~(\sigma)$}     &
{$R$}     \\
{} &
{}                   &
{Rate}                   &
{}                   &
{(0.5--10.0 keV)}                   &
{}                   &
{}                   &
{}     &
{}     &
{}     \\
{(1)}         &
{(2)}         &
{(3)}         &
{(4)}         &
{(5)}         &
{(6)}         &
{(7)}         &
{(8)}         &
{(9)}         &
{(10)}         \\
\hline
            J$0401-0540$&$-26.0$&$         9.29_{-0.49}^{+0.52}$& 72.06& 44.98& 8.96&30.88&$  -1.57$&$   0.02( 0.11)$&$<  0.7$\\
            J$0850+4451$&$-25.1$&$         0.04_{-0.02}^{+0.04}$&  0.32& 42.43& 2.27&30.24&$  -2.25$&$  -0.74(-3.75)$&$<  1.2$\\
            J$0959+3449$&$-25.0$&$         5.29_{-0.24}^{+0.25}$& 42.02& 44.92& 1.88&30.39&$  -1.40$&$   0.12( 0.63)$&$<  1.6$\\
            J$1500+1210$&$-27.2$&$         5.77_{-0.37}^{+0.40}$& 37.77& 45.19&12.30&31.30&$  -1.73$&$  -0.08(-0.56)$&$   0.7$\\
            J$1531+2420$&$-25.9$&$        15.44_{-0.59}^{+0.62}$& 95.22& 45.25& 7.22&30.88&$  -1.49$&$   0.10( 0.52)$&$   6.2$\\
            J$1609+0824$&$-24.8$&$         9.11_{-0.33}^{+0.34}$& 74.46& 45.22& 2.70&30.58&$  -1.37$&$   0.19( 0.94)$&$<  1.1$\\
            J$2125-0813$&$-25.8$&$         8.89_{-0.17}^{+0.17}$& 58.72& 45.10& 3.78&30.59&$  -1.46$&$   0.09( 0.46)$&$   3.1$\\
            J$2148+0010$&$-25.5$&$         1.03_{-0.12}^{+0.13}$&  6.02& 44.42& 4.90&30.77&$  -1.88$&$  -0.31(-1.55)$&$<  0.5$\\
\hline
\end{tabular}

Note. --- Col. (1): abbreviated SDSS name.
Col. (2): absolute $i$-band magnitude from SDSS.
Col. (3): count rate in the 0.5--8.0 keV band in units of $10^{-2}$~s$^{-1}$.
Col. (4): flux density at rest-frame 2 keV in units of $10^{-32}$ \mflux.
Col. (5): logarithm of the X-ray luminosity in the rest-frame 0.5--10.0 keV band corrected for Galactic absorption.
Col. (6): flux density at rest-frame 2500 \AA\ in units of $10^{-27}$ \mflux.
Col. (7): logarithm of the monochromatic luminosity at rest-frame 2500 \AA.
Col. (8): X-ray-to-optical power-law slope, defined as $\alpha_{\rm OX}=-0.3838\log(F_{\rm 2500~\AA}/F_{\rm 2~keV})$. The flux density is measured per unit frequency.
Col. (9): the difference between the measured and expected $\alpha_{\rm OX}$. $\Delta\alpha_{\rm OX}=\alpha_{\rm OX(measured)}-\alpha_{\rm OX(expected)}$. The statistical significance of this difference, $\sigma$, is measured in units of the rms of the expected $\alpha_{\rm OX}$ as given in Table 5 of \citet{Steffen2006}.
Col. (10): radio-loudness parameter (see Section 5.1).
\label{tbl-aox}
\end{minipage}
\end{table*}

\end{document}